\documentclass[prl,a4paper,10pt,twocolumn,floatfix,superscriptaddress,amsmath,amsfonts,amssymb,preprintnumbers,citeautoscript]{revtex4}
\pdfoutput=1
\usepackage[T1]{fontenc}\usepackage[latin1]{inputenc}
\usepackage{dcolumn,graphicx,color,booktabs,microtype,afterpage} 
\usepackage{sidecap}
\usepackage{times}
\usepackage[colorlinks,plainpages=false,linkcolor=blue,urlcolor=blue,citecolor=blue,pdfpagemode=UseNone,pdfstartview=FitBH]{hyperref}

\usepackage{amssymb}
\usepackage{amsmath}
\usepackage{graphicx}

\newcommand{\ybsio}{Yb$_2$Si$_2$O$_7$}
\newcommand{\hctwo}{$H_{c2}$}
\newcommand{\hcone}{$H_{c1}$}
\newcommand{\hu}{$H_m$}

\newcommand{\Seff}{$S_{\text{eff}}$}
\begin{document}
%
%
%
%
\title
{A Novel Strongly Spin-Orbit Coupled Quantum Dimer Magnet: \ybsio}
\author{Gavin Hester}
\email{Gavin.Hester@colostate.edu}
\affiliation{Department of Physics, Colorado State University, 200 W. Lake St., Fort Collins, CO 80523-1875, USA}
\author{H. S. Nair}
\affiliation{Department of Physics, Colorado State University, 200 W. Lake St., Fort Collins, CO 80523-1875, USA}
\author{T. Reeder}
\affiliation{Department of Physics, Colorado State University, 200 W. Lake St., Fort Collins, CO 80523-1875, USA}
\author{D. R. Yahne}
\affiliation{Department of Physics, Colorado State University, 200 W. Lake St., Fort Collins, CO 80523-1875, USA}
\author{T. N. DeLazzer}
\affiliation{Department of Physics, Colorado State University, 200 W. Lake St., Fort Collins, CO 80523-1875, USA}
\author{L. Berges}
\affiliation{Institut Quantique and D\'epartement de Physique, Universit\'e de Sherbrooke, 2500 boulevard de l'Universit\'e, Sherbrooke, Qu\'ebec J1K 2R1, Canada}
\author{D. Ziat}
\affiliation{Institut Quantique and D\'epartement de Physique, Universit\'e de Sherbrooke, 2500 boulevard de l'Universit\'e, Sherbrooke, Qu\'ebec J1K 2R1, Canada}
\author{J. R. Neilson}
\affiliation{Department of Chemistry, Colorado State University, 200 W. Lake St., Fort Collins, CO 80523-1872, USA}
\author{A. A. Aczel}
\affiliation{Neutron Scattering Division, Oak Ridge National Laboratory, Oak Ridge, TN 37831, USA}
\author{G. Sala}
\affiliation{Neutron Scattering Division, Oak Ridge National Laboratory, Oak Ridge, TN 37831, USA}
\author{J. A. Quilliam}
\email{Jeffrey.Quilliam@USherbrooke.ca}
\affiliation{Institut Quantique and D\'epartement de Physique, Universit\'e de Sherbrooke, 2500 boulevard de l'Universit\'e, Sherbrooke, Qu\'ebec J1K 2R1, Canada}
\author{K. A. Ross}
\email{Kate.Ross@colostate.edu}
\affiliation{Department of Physics, Colorado State University, 200 W. Lake St., Fort Collins, CO 80523-1875, USA}
\affiliation{Quantum Materials Program, Canadian Institute for Advanced Research (CIFAR), Toronto, Ontario M5G 1Z8, Canada}
\date{\today}
%
%
%
%
\begin{abstract}
The quantum dimer magnet (QDM) is the canonical example of quantum magnetism. The QDM state consists of entangled nearest-neighbor spin dimers and often exhibits a field-induced triplon Bose-Einstein condensate (BEC) phase. We report on a new QDM in the strongly spin-orbit coupled, distorted honeycomb-lattice material \ybsio. Our single crystal neutron scattering, specific heat, and ultrasound velocity measurements reveal a gapped singlet ground state at zero field with sharp, dispersive excitations. We find a field-induced magnetically ordered phase reminiscent of a BEC phase, with exceptionally low critical fields of $H_{c1} \sim 0.4$~T and $H_{c2} \sim 1.4$~T. Using inelastic neutron scattering in an applied magnetic field we observe a Goldstone mode (gapless to within $\delta E$ = 0.037 meV) that persists throughout the entire field-induced magnetically ordered phase, suggestive of the spontaneous breaking of U(1) symmetry expected for a triplon BEC. However, in contrast to other well-known cases of this phase, the high-field ($\mu$$_0$$H\geq1.2$T) part of the phase diagram in \ybsio\ is interrupted by an unusual regime signaled by a change in the field dependence of the ultrasound velocity and magnetization, as well as the disappearance of a sharp anomaly in the specific heat. These measurements raise the question of how anisotropy in strongly spin-orbit coupled materials modifies the field induced phases of QDMs. 
\end{abstract}

\maketitle
%
%
%
%
\indent
Quantum dimer magnets (QDMs) represent the simplest cases of quantum magnetism, where entanglement is a required ingredient for even a qualitative understanding of the phase. In a QDM, entangled pairs of spins form $S_{tot}=$ 0 dimers and result in a non-magnetic ground state. The excited states of these entangled spins can be treated as bosons, called triplons, which can undergo Bose-Einstein condensation (BEC) as their density is tuned by an applied magnetic field. This BEC state is a magnetic field-induced long range ordered phase, which occupies a symmetric ``dome'' in the field vs. temperature phase diagram with two temperature-dependent critical fields, $H_{c1}(T)$ and $H_{c2}(T)$. The vast majority of the previously studied QDMs are based on 3d transition metal ions with ``bare'' (spin-only) $S=1/2$ or $S=1$ angular momentum, resulting in simple Heisenberg or XXZ spin interaction Hamiltonians, and high critical fields set by the relatively high energy scale of exchange interactions \cite{Zapf2014, Zapf2006, Jaime2004, Samulon2008, Aczel2009, aczel2009bose}.

Lanthanide-based magnetic materials with spin-orbit coupled pseudo-spin 1/2 (\Seff\ = 1/2) angular momenta can also exhibit quantum phases, and these are often directly analogous to their traditional 3d transition metal ion counterparts. However, entirely new phases are possible due to the anisotropic exchange in these materials \cite{hermele2004pyrochlore,onoda2010quantum,ross2011,gingras2014quantum,Kitaev2006,jackeli2009mott}. In the lanthanide series, Yb$^{3+}$ has been of particular interest as it can generically host interactions leading to quantum fluctuations irrespective of the Crystal Electric Field (CEF) ground state doublet composition \cite{Rau2018}. Indeed, various quantum phases have been discovered in Yb-based systems \cite{Hallas2016, Wu2016, Wu2019, ono1970rare, yb4as3,Rau2016}. Recently, a random valence bond state in YbMgGaO$_4$ was proposed \cite{Kimchi2018}. However, a notable absence in the growing lineup of Yb quantum materials is a material exhibiting a QDM with a field-induced BEC state. The opportunity to study such a material could lead to the observation of new phases describable by theories of interacting bosons, as well as new types of quantum phase transitions.

As a previously studied example, the metallic material YbAl$_3$C$_3$ was shown to host Yb dimerization and triplet excitations \cite{Ochiai2007, Kato2008}. However, an unusual field-induced ordered state was observed whose onset temperature far exceeds the spin gap energy \cite{Khalyavin2013}, suggesting that it is not directly related to the singlet-triplet excitation (unlike a field-induced BEC phase). Additionally, YbAl$_3$C$_3$ shows field-induced \emph{disordered} regimes that have yet to be fully understood, particularly in the context of the additional Kondo and RKKY interactions involving the conduction electrons in this material \cite{Nakanishi2012, Hara2012, Kittaka2013}. This material demonstrates that quantum dimerization is possible in lanthanide-based magnetic materials, but does not always lead to a field-induced BEC phase. Naively, one might not expect a highly spin-orbit coupled material to exhibit BEC, which requires the exchange Hamiltonian to be at least U(1) symmetric (i.e., XXZ type interactions). Although recent work has demonstrated that for ideal, edge-sharing octahedral environments, Heisenberg exchange is indeed expected to dominate in Yb materials \cite{Rau2018}, such high exchange symmetry is not \emph{a priori} expected for non-ideal local environments. However, a recent example of high exchange symmetry for Yb$^{3+}$ in a non-ideal crystal field environment has been discovered in the Tomonaga-Luttinger liquid YbAlO$_3$ \cite{Wu2019}, suggesting that it may be more common than expected. Yet even with dominant Heisenberg interactions, smaller anisotropic terms should still be relevant which, in the case of a QDM, would be expected to modify the field-induced phases. Furthermore, Yb-based QDMs should provide a convenient testing ground for field-induced BEC physics due to reduced exchange energy compared to materials based on 3d transition metals. This leads to lower critical fields, which can be accessed by continuous field magnets, thus enabling experimental techniques such as inelastic neutron scattering (INS) to be brought to bear on the full phase diagram. This is the case for \ybsio, as we show here. 

\begin{figure}[!t]
\includegraphics[width = \columnwidth]{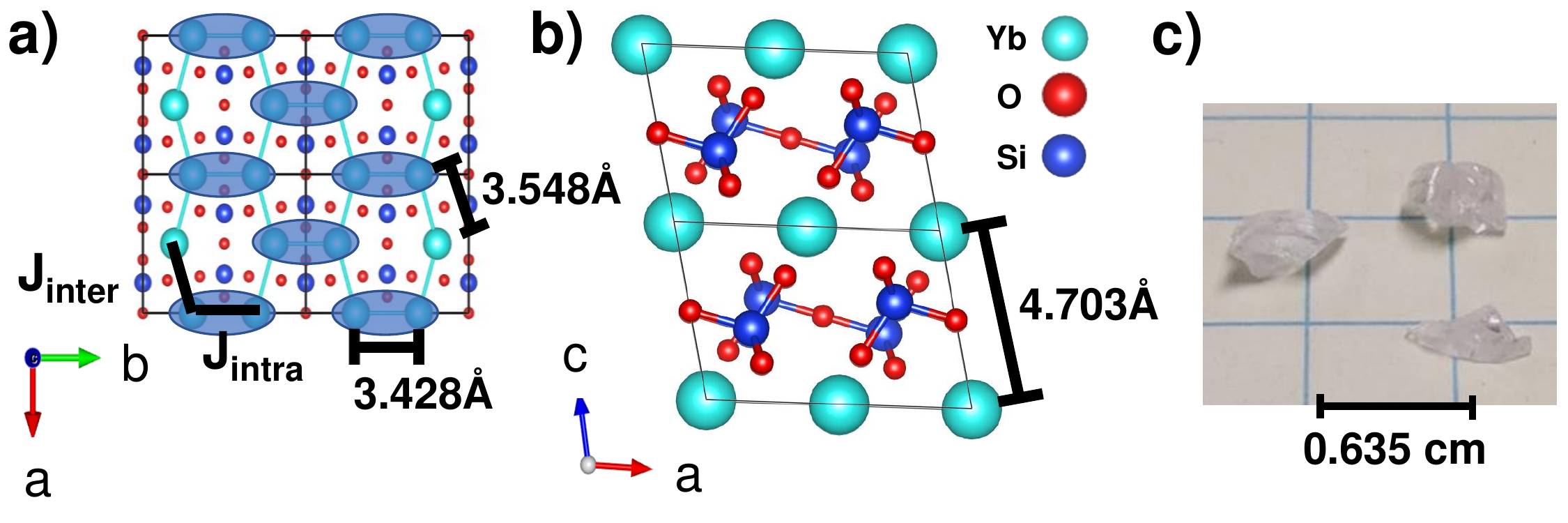}
\caption{ a) Crystal structure of \ybsio\, viewed along the $c$-axis, where Yb atoms are light green and form a distorted honeycomb lattice, Si atoms are blue, and O atoms are red \cite{Vesta}. Intradimer and interdimer bond lengths are shown (3\% anisotropy), and $J_{\text{intra}}$ and $J_{\text{inter}}$ exchange tensors are labeled. The blue ovals indicate the probable location of the dimers. b) Crystal structure viewed along the $b$-axis, showing the separation of the layers of Yb honeycombs. c) Characteristic crystals obtained from breaking the crystal boule. The crystals are clear and colorless.} \label{fig_crystStruc}
\end{figure}
\ybsio\ (monoclinic space group C$2/m$, room temperature lattice parameters of $a = $6.7714(9)\AA, $b = $ 8.8394(2)\AA, $c = $4.6896(5)\AA, $\beta$ = 101.984(9)$^\circ$ \cite{SI}) was previously studied in the context of polymorphism in the RE$_2$Si$_2$O$_7$ (rare earth pyrosilicate) series \cite{Felsche1970, smolin1970crystal}, but its magnetic properties have not been reported. \ybsio\ has only one reported polymorph, known as the C-type pyrosilicate (Fig.~\ref{fig_crystStruc}). The single crystal samples of \ybsio\ used in this study were grown via the optical floating zone method \cite{SI, Nair2019}. Our growths have resulted in clear, colorless multi-crystal boules which are then broken into smaller single crystal pieces as shown in Fig.~\ref{fig_crystStruc}c. 

Magnetization was measured using a MPMS XL Quantum Design SQUID magnetometer at $T=$ 1.8 K along the $a^*$, $b$, and $c$ directions. Field and temperature-dependent specific heat was measured down to 50 mK using the quasi-adiabatic heat pulse method in a Quantum Design Dynacool PPMS with a dilution refrigerator insert at Colorado State University, as well as a home-built dilution refrigerator at Universit\'e de Sherbrooke. Lu$_2$Si$_2$O$_7$ was also measured as a non-magnetic analog. Ultrasound velocity experiments were performed down to 50 mK using a pulsed, time-of-flight interferometer. 30 MHz transducers were glued to parallel surfaces so as to propagate longitudinally polarized sound waves along the $c^\ast$-axis. The absolute velocity of the quasi-longitudinal mode studied here was approximately 3000 m/s and relative changes in velocity ($\Delta v/v$) were measured with high precision using a phase-lock loop. Powder neutron diffraction data was collected on BT1 at the NIST Center for Neutron Research with incident wavelength $\lambda = 2.0787$ \AA\ and 60 arcminute collimation. Synchrotron x-ray diffraction (SXRD) data were recorded at $T=295$~K at beamline 11 BM ($\lambda$ = 0.41418~{\AA}) at the Advanced Photon Source, Argonne National Laboratory. Time-of-flight INS experiments were performed at the Cold Neutron Chopper Spectrometer (CNCS) at the Spallation Neutron Source, Oak Ridge National Laboratory (ORNL). These INS data were collected using $E_i$~=~1.55~meV neutrons in the ``high flux'' chopper setting mode, producing an energy resolution of $\delta E$ = 0.037 meV at the elastic line \cite{ehlers2016cold}, and were analyzed using the DAVE software package \cite{DAVE}. A neutron diffraction measurement using $E_i = $ 14.7 meV neutrons was performed using the Fixed-Incident Energy Triple-Axis Spectrometer (FIE-TAX) on the HB-1A beamline at the High Flux Isotope Reactor at Oak Ridge National Laboratory, using collimator settings of 40' - 40' - 40' - 80'.

Rietveld analysis of the SXRD data \cite{SI} confirms the previously reported crystal structure. Analysis of the zero field, high-temperature, magnetic specific heat of \ybsio\ confirms that a low energy \Seff=1/2 picture applies at temperatures well below $\sim 100$ K \cite{SI}. The saturation magnetization at $T=1.8$ K along three crystal directions gives the approximate $g$-values of $g_{a*}~=~3.2$, $g_{b}~=~2.0,$ and $g_{c}~=~4.8$. %
\begin{figure}[!t]
\includegraphics[width = \columnwidth]{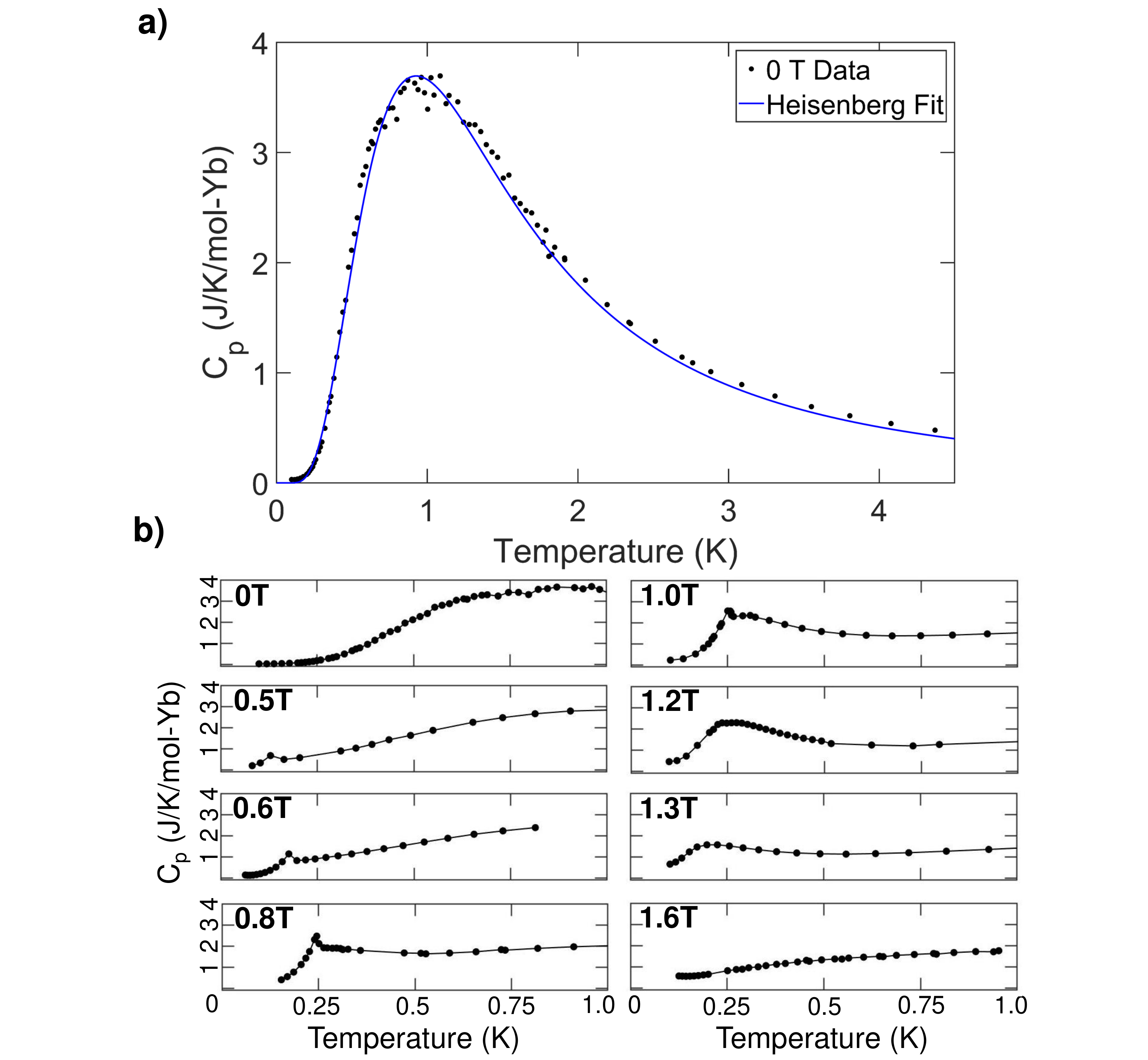}
\caption{a) Zero-field specific heat and fit to a dispersive 4-level Schottky anomaly, using Heisenberg exchange for inter- and intra-dimer interactions ($J_{\text{intra}}$ = 0.236(4) meV, $J_{\text{inter}}$ = 0.06(2) meV). b) Specific heat of \ybsio\ at increasing fields with $H||c$. A sharp anomaly is visible at 0.5 T ($>$\hcone), which corresponds to a field-induced magnetically ordered state. The transition temperature maps out a dome as a function of field, but the sharp anomaly is replaced by a broad anomaly above $\sim1.2$ T (\hu), which moves to lower temperatures with increasing field. Above \hctwo (1.4 T), the broad anomaly shifts to higher temperatures with increasing field, consistent with field polarized paramagnetism. \label{fig_specificheat}}
\end{figure}

\begin{figure*}[ht]
\includegraphics[width = 2\columnwidth]{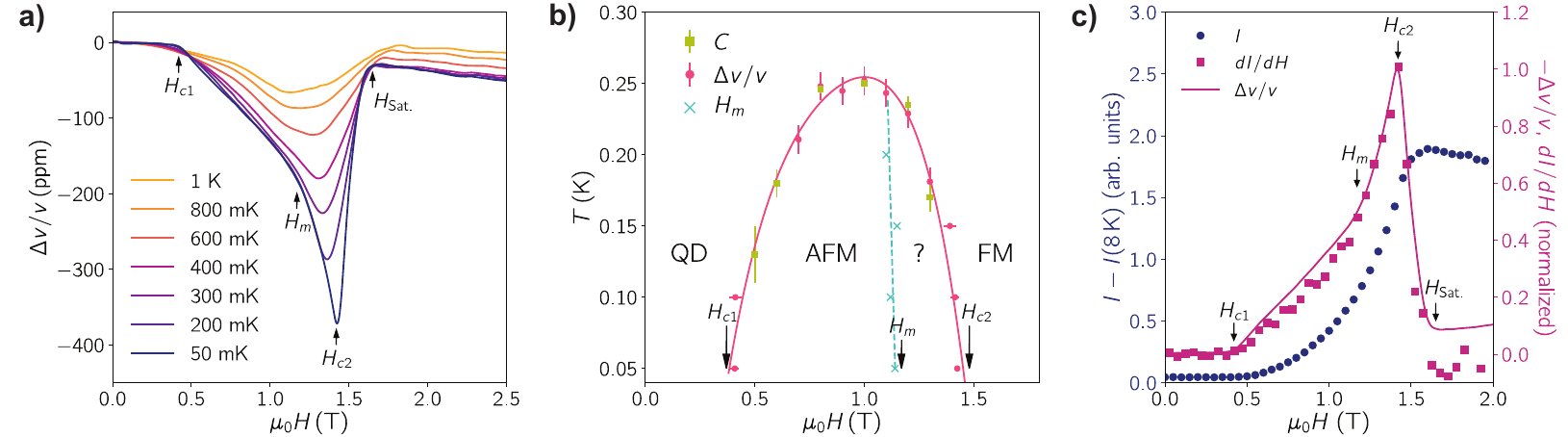}
\caption{a) Ultrasound velocity with longitudinally polarized sound waves along the $c^\ast$-axis. b) $H$ vs. $T$ phase diagram for \ybsio\, with the points on the phase boundary determined by ultrasound velocity (pink circles and blue crosses) and specific heat (yellow squares). The field was applied along the $c$-axis (specific heat) and $c^\ast$-axis (ultrasound). c) Evolution of the (2,0,0) magnetic Bragg peak intensity (blue) versus field, $I(H)$, which is proportional to the square of the net magnetization. Additionally the derivative of the (2,0,0) magnetic Bragg peak intensity (square symbols) and the inverse of the ultrasound velocity data (solid line) are overlaid, showing agreement between these two measurements.} \label{fig_uv}
\end{figure*}

The zero-field specific heat shown in Fig.~\ref{fig_specificheat}a displays a broad feature peaked at $\sim 1$ K, which can be fit to a dispersive four level Schottky anomaly form, consistent with an interacting spin dimer ground state. We used an approximation of an interacting triplon model to fit the zero-field specific heat \cite{SI}, enforcing Heisenberg interactions. The fit yielded the parameters $J_{\text{intra}}$ = 0.236(4) meV and $J_{\text{inter}}$ = 0.06(2) meV. These parameters are similar to those extracted from fitting the field polarized spin wave spectrum; $J_{\text{intra}}$ = 0.217(3) meV and $J_{\text{inter}}$ = 0.089(1) meV \cite{SI}. The adequacy of Heisenberg interactions for reproducing both the zero field $C_p$ and field-polarized INS data measurements suggests that \ybsio\ is another case in which Yb$^{3+}$ interactions are unexpectedly predominantly isotropic. The entropy change through this low temperature Schottky anomaly (0.05 to 2 K), reaches the expected $R$ln2 per Yb \cite{SI}, indicating that \ybsio\ does not undergo a magnetic ordering transition at lower temperatures, and thus remains quantum disordered down to $T=0$ K. This is further confirmed by the lack of magnetic Bragg peaks at 50 mK, as determined by both single crystal (Fig.~\ref{fig_uv}c) and powder neutron diffraction measurements \cite{SI}.

The field-dependence ($H || c$) of the specific heat is shown in Fig.~\ref{fig_specificheat}b. At $H= 0.5$ T, a sharp anomaly appears at $T = 0.13$ K, which we have confirmed by neutron scattering to coincide with a transition to long range magnetic order via the appearance of magnetic Bragg peaks. With increasing field, the transition temperature maps out a ``dome'' in the $H$ vs. $T$ phase diagram as expected for a BEC phase. As the field is increased further (0.8T), a broad feature emerges, which eventually becomes the dominant feature above $H_m = 1.2$ T. The maximum of this broad feature then continues to trace out the high field region of the dome, with the temperature of the maximum decreasing with increasing field. At 1.6 T, the maximum of the broad feature is again increasing in temperature with increasing field as expected for a field-polarized paramagnetic regime.

Isothermal field scans of variations in sound velocity are shown in Fig.~\ref{fig_uv}a for various temperatures. At the lowest temperatures ($T~=~50$~mK) the sound velocity is largely field independent until $H_{c1} \simeq 0.4$ T, where $\Delta v/v$ begins decreasing with field. At $H_{c2} \simeq 1.4$ T, $\Delta v(H)$ reaches a minimum, before returning sharply to roughly the zero field value in the field polarized limit. In addition to the two expected critical fields, $H_{c1}$ and $H_{c2}$, the sound velocity also exhibits a significant change in slope at roughly $H_m = 1.2$ T, suggesting the presence of an additional phase, as indicated in Fig.~\ref{fig_uv}b. Aside from the sharp change of slope at $H_m$, our sound velocity measurements resemble those performed on another quantum dimer magnet, Sr$_3$Cr$_2$O$_8$~\cite{Wang2016}. In contrast, sound velocity measurements on NiCl$_2$-4SC(NH$_2$)$_2$~\cite{Chiatti2008} show sharper dips at both $H_{c1}$ and $H_{c2}$, which are attributed to coupling between the ultrasound velocity and antiferromagnetic fluctuations. 

As the temperature is raised, the overall variations in sound velocity become much smaller in magnitude and the sharp features are smoothed out, hence we use temperature scans of sound velocity (see Supplemental Information~\cite{SI}), which show small but fairly sharp anomalies, to establish the phase boundaries of the antiferromagnetic dome at higher temperatures. These boundaries are entirely consistent with the specific heat measurements.

The dome of field-induced order mapped out by the specific heat and ultrasound velocity data (Fig.~\ref{fig_uv}b) is similar to the BEC phase of traditional QDMs, but there is an important difference: the dome in \ybsio\ is highly asymmetric, with an unusual regime in the high field part of the phase ($H > H_m$). Asymmetry of the dome can sometimes be attributed to quantum fluctuations in the proximity of \hcone\, which is expected when \hcone/(\hctwo-\hcone) is small.  However, in \ybsio\ this number is 0.4, which is twice as large as the well-known case of dome asymmetry in DTN \cite{Kohama2011}.  Further, this effect does not explain the high field phase above $H_m$. This unusual regime may be due to non-U(1) symmetric terms in the \Seff=1/2 low energy effective Hamiltonian for \ybsio. However, the strength of any anisotropic exchange is limited by our observation of a Goldstone-like mode (gapless to within $\delta E = 0.037$ meV) via INS, discussed below. 

Fig.~\ref{fig_uv}c shows the field dependence of neutron diffraction (measured on FIE-TAX) at the (2,0,0) zone center. This reflection is only sensitive to the square of the net magnetization ($m_z^2$) that arises due to canting towards the field direction\, rather than any AFM components of the magnetic structure. The onset of magnetic order and growth of the net magnetization is confirmed above \hcone\ through the observation of increasing magnetic Bragg peak intensity. The intensity of the (2,0,0) peak shows an approximately quadratic increase, with a sudden change in the second derivative occurring at approximately $H_m$. Additionally, Fig. ~\ref{fig_uv}c shows a comparison of the first derivative of the (2,0,0) Bragg peak intensity at 50~mK and the negative of the relative ultrasound velocity at 100~mK, which are consistent (though this level of agreement is somewhat unexpected following a standard theoretical treatment, see \cite{SI}).

\begin{figure}[!t]
\includegraphics[width = \columnwidth]{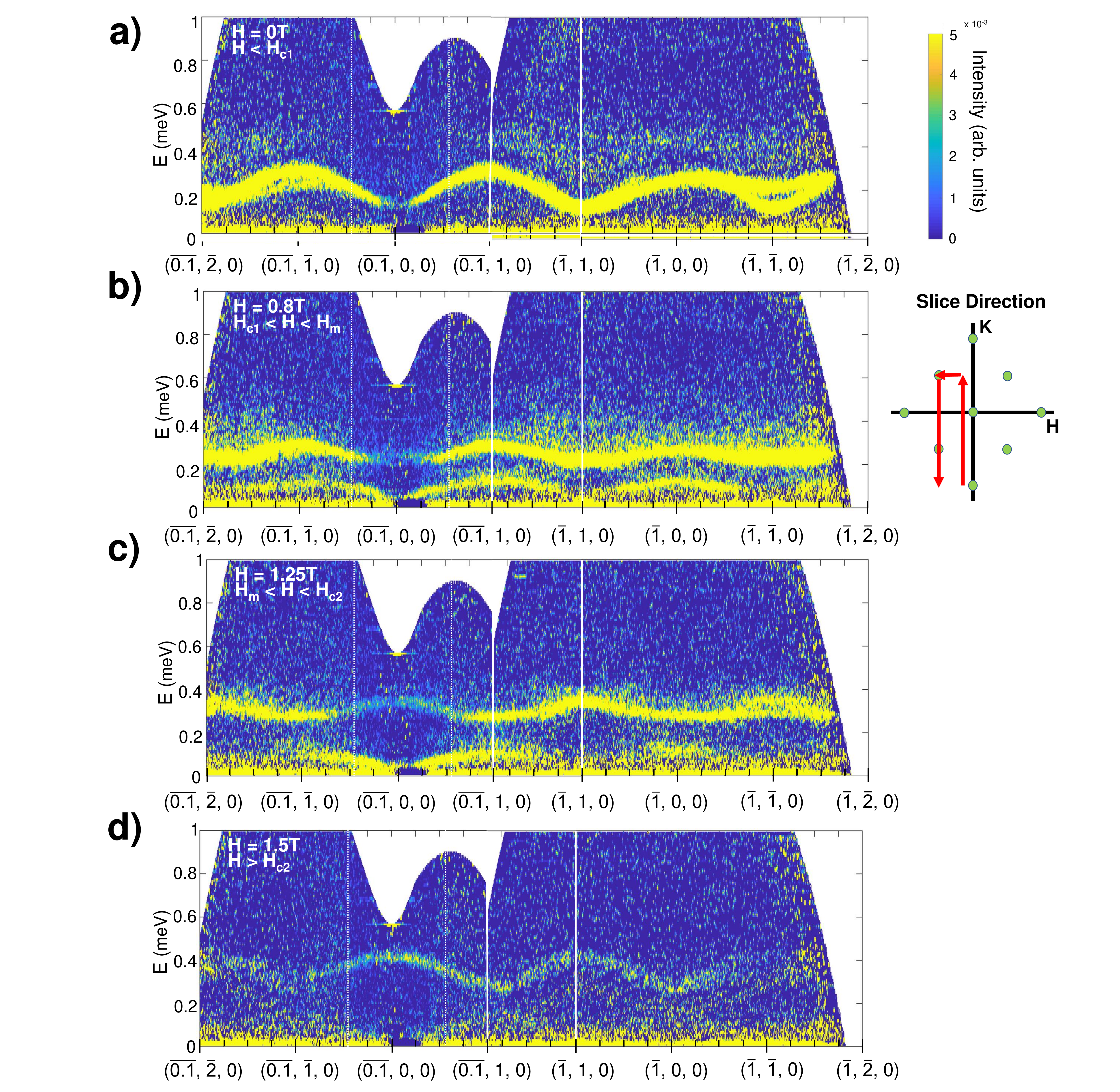}
\caption{INS data at $T = 50$ mK for four representative field strengths ($H || c$). The path shown includes the reciprocal lattice directions [-0.1K0], [H10], and [-1K0] as shown schematically to the right of the figure. All slices shown are integrated $\pm$ 0.1 r.l.u. in the perpendicular direction. At zero field (panel a), two bands are visible near (-1,1,0) and (-0.1, -1, 0) due to a misaligned grain in the sample \cite{SI}. These are actually due to the same excitation which is identified as the $\psi_{1,0}$ state. Between \hcone\ and \hctwo\ (panels b and c), a Goldstone mode appears which is gapless at zone centers to within the energy resolution of the instrument, $\delta E$ = 0.037 meV. Above \hctwo\ (panel d) the intensity of the excitation drops dramatically due to the system entering a field-polarized paramagnet state. \label{fig_neutrons}}
\end{figure}
INS data provides evidence of the spontaneous breaking of an approximately continuous symmetry for fields between \hcone\ and \hctwo. Fig.~\ref{fig_neutrons} shows the INS spectra of \ybsio\ at $T=50$~mK for representative applied fields along the $c$-axis. In a QDM with Heisenberg exchange, the three excited dimer states are triply degenerate (forming a triplet with $S_{tot}=1$, and $S_z = -1, 0,$ and 1), and are then Zeeman split by the applied magnetic field. With finite interdimer exchange the resulting triplons are mobile, and the excited states become dispersive. For \ybsio\ below \hcone\, a resolution-limited single excited dispersive branch (bandwidth of 0.167(1) meV, and a gap of 0.1162(4) meV) is visible. The apparent secondary branch observed around $(\overline{0.1},\overline{1},0)$ and (\={1},\={1},0) is due to a minority crystal grain. The energy of the observed excitation does not change for $H < $\hcone\, as shown in the supplemental information \cite{SI}, signifying that the angular momentum projection along the magnetic field is zero (i.e., S$_{tot}$ = 1, $S_z$ = 0, which we call $\psi_{1,0}$). The absence of apparent $S_{tot} = 1$, $S_z \pm 1$ modes (hereafter labeled as $\psi_{1,\pm 1}$) at most field strengths below \hcone\ indicates that the neutron scattering transition matrix elements from the ground state to $\psi_{1,\pm 1}$ are small compared to that for $\psi_{1,0}$. However, $\psi_{1,\pm 1}$ are discernible with very weak intensity at fields near \hcone\, indicating the transition matrix elements are non-zero \cite{SI}. Above \hcone, a new low energy excitation appears, which is gapless at the magnetic zone centers to within the energy resolution of the instrument ($\delta E = $ 0.037 meV). This Goldstone mode implies spontaneous breaking of an approximate U(1) symmetry in the plane perpendicular to the applied magnetic field (the $a^*$-$b$ plane), suggestive of the BEC transition observed in traditional QDMs \cite{Zapf2014, Giamarchi2008}. Additionally we note that the energy resolution is $\sim$16\% of our estimated $J_{\text{intra}}$, thus this measurement of the Goldstone mode actually allows for a potentially sizable anisotropic exchange contribution. Furthermore, the presence of a distinguishable region of hte field-induced phase (between $H_m$ and \hctwo) is not expected for simple Heisenberg or XXZ exchange. We find that in this field region the Goldstone mode persists, despite the lack of evidence for spontaneous symmetry breaking in $C_p(T)$ (i.e. a sharp anomaly is absent). However, the broad $C_p(T)$ feature does move to lower temperature as the field is further increased in this field region, tracing out the high-field side of the dome phase boundary. Above \hctwo\ all of the excitations become fully gapped and the broad feature in $C_p$ moves to higher temperature with increasing field, consistent with a field-polarized paramagnet. In the field polarized regime, the inelastic intensity is greatly reduced due to the development of strong magnetic Bragg peaks at the elastic line, as expected based on the sum rule for magnetic neutron scattering.


Recently, rare-earth materials have been identified as potential hosts of Kitaev exchange in honeycomb materials \cite{GangChen}. In light of this, it is important to note that \ybsio\ is structurally similar to the famous Kitaev material Na$_2$IrO$_3$ \cite{Chun2015}, as they share the same space group and Wyckoff position of the magnetic species. Therefore, Kitaev exchange is allowed by symmetry in \ybsio. If Kitaev exchange were dominant in \ybsio\, it could lead to a quantum spin liquid ground state \cite{Kitaev2006}. Interestingly, the presence of a Goldstone mode does \emph{not} rule out such anisotropic Kitaev exchange due to the ``hidden'' SU(2) symmetries found within the extended Kitaev-Heisenberg model \cite{Jackeli2013, Chaloupka2015}. However, our fits to field polarized INS data are well-approximated by Heisenberg interactions, so Kitaev interactions are unlikely to be dominant in this material.


In summary, the strongly spin-orbit coupled material \ybsio\ realizes a QDM ground state with magnetic field-induced order reminiscent of a BEC phase. However, this ordered phase exhibits unusual characteristics at the high field part of the dome, including an abrupt change in the field dependence of the magnetization and sound velocity, and the loss of a sharp anomaly in the specific heat. The presence of a Goldstone mode throughout the full field-induced ordered state suggests dominant Heisenberg or XXZ exchange interactions, and the former is confirmed by fits to field polarized INS data and the zero field specific heat. However, the observation of the unusual regime between $H_m$ and \hctwo\ may imply that additional anisotropic interactions are necessary in order to fully describe the field induced phases of this novel quantum magnet. \ybsio\ provides the first example of a Yb$^{3+}$-based QDM with a possible field-induced BEC phase, adding this canonical example of quantum magnetism to the roster of quantum phases exhibited by materials based on this versatile ion.

\begin{acknowledgements}
This research was supported by the National Science Foundation Agreement No. DMR-1611217. JQ acknowledges technical support from M. Castonguay and S. Fortier, informative conversations with G. Quirion, C. Bourbonnais and  I. Garate and funding from NSERC. The authors acknowledge the assistance of Aaron Glock and Antony Sikorski in the sample synthesis, as well as Craig Brown for his assistance with the BT1 neutron powder diffraction experiment. A portion of this work used resources at the Spallation Neutron Source and High Flux Isotope Reactor, which are DOE Office of Science User Facilities operated by Oak Ridge National Laboratory. The authors also acknowledges the support of the National Institute of Standards and Technology, US Department of Commerce in providing some of the neutron research facilities used in this work.
\end{acknowledgements}

%
%

\end{document}


\preprint{APS/123-QED}
%
%
\title
{Supplemental Information for ``A Novel Strongly Spin-Orbit Coupled Quantum Dimer Magnet: \ybsio''}
\author{Gavin Hester}
\affiliation{Department of Physics, Colorado State University, 200 W. Lake St., Fort Collins, CO 80523-1875, USA}
\author{H. S. Nair}
\affiliation{Department of Physics, Colorado State University, 200 W. Lake St., Fort Collins, CO 80523-1875, USA}
\author{T. Reeder}
\affiliation{Department of Physics, Colorado State University, 200 W. Lake St., Fort Collins, CO 80523-1875, USA}
\author{D. R. Yahne}
\affiliation{Department of Physics, Colorado State University, 200 W. Lake St., Fort Collins, CO 80523-1875, USA}
\author{T. N. DeLazzer}
\affiliation{Department of Physics, Colorado State University, 200 W. Lake St., Fort Collins, CO 80523-1875, USA}
\author{L. Berges}
\affiliation{Institut Quantique and D\'epartement de Physique, Universit\'e de Sherbrooke, 2500 boulevard de l'Universit\'e, Sherbrooke, Qu\'ebec J1K 2R1, Canada}
\author{D. Ziat}
\affiliation{Institut Quantique and D\'epartement de Physique, Universit\'e de Sherbrooke, 2500 boulevard de l'Universit\'e, Sherbrooke, Qu\'ebec J1K 2R1, Canada}
\author{J. R. Neilson}
\affiliation{Department of Chemistry, Colorado State University, 200 W. Lake St., Fort Collins, CO 80523-1875, USA}
\author{A. A. Aczel}
\affiliation{Neutron Scattering Division, Oak Ridge National Laboratory, Oak Ridge, TN 37831, USA}
\author{G. Sala}
\affiliation{Neutron Scattering Division, Oak Ridge National Laboratory, Oak Ridge, TN 37831, USA}
\author{J. A. Quilliam}
\affiliation{Institut Quantique and D\'epartement de Physique, Universit\'e de Sherbrooke, 2500 boulevard de l'Universit\'e, Sherbrooke, Qu\'ebec J1K 2R1, Canada}
\author{K. A. Ross}
\affiliation{Department of Physics, Colorado State University, 200 W. Lake St., Fort Collins, CO 80523-1875, USA}
\affiliation{Quantum Materials Program, Canadian Institute for Advanced Research (CIFAR), Toronto, Ontario M5G 1Z8, Canada}
\date{\today}
\maketitle

\section{Sample Preparation}
Polycrystalline \ybsio\ was synthesized by combining stoichiometric amounts of Yb$_2$O$_3$ and SiO$_2$, pressing under hydrostatic pressure of $\sim$480 kPa, and heating 4-5 times at 1350$^\circ$C for 48 hours, with regrinding between heatings to promote reaction, until phase purity was achieved (as confirmed by powder x-ray diffraction in air). Sintered cylindrical rods with diameter of 8 mm were prepared from these powders for optical floating zone crystal growth. A Crystal Systems furnace (FZ-T-10000-H-VIII-VPO-PC) was used for the crystal growth. Multiple growths were performed to optimize the parameters. The most successful growths were performed with 1.5 kW lamps (70-73\%\ power), with a growth rate of 3-5 mm/hr, under an atmosphere of flowing oxygen (1-2L / min), with a counterrotation of the upper and lower rods of 20 rpm. Every growth resulted in cracked boules, which upon further study by Laue x-ray diffraction, were found to be multi-crystalline. The boules were broken into separate single crystals (typical size approximately 3 $\times$ 3 $\times$ 2 mm$^3$) which were clear and colorless (see Fig. 1c of main text). 

%

\section{Crystal Electric Field considerations}

The low point group symmetry of Yb$^{3+}$ in \ybsio\ (C$_2$) leads to nine independent Steven's parameters in a crystal field Hamiltonian \cite{mcphase_manual}. Determining these experimentally, for example by an inelastic neutron scattering measurement of the single ion energy levels, is an underdetermined problem, since such an experiment only gives access to three transitions (between the four Kramers doublets). Thus, the CEF ground state for \ybsio\ is experimentally unknown. However, our observations do restrict some of the properties of the CEF ground state. Perhaps most significantly, we find that the $\psi_{1,\pm1}$ modes are not easily visible via inelastic neutron scattering at $E_i$ = 1.55 meV. This can be explained if the CEF ground state doublet for Yb$^{3+}$ does not have a significant matrix element for transitions induced by $J_+$ or $J_-$. For example, a CEF ground state doublet that is composed primarily of a single $|J_z\rangle$ eigenstate (except $|J_z\rangle = \pm 1/2$), will have vanishingly small matrix elements to the excited $\psi_{1,\pm1}$ states, as discussed below.

Assuming XXZ symmetry for the intradimer interactions, the dimer eigenstates are given by:

\begin{align*} 
&|\psi_{0,0}\rangle = \frac{1}{\sqrt{2}}(|\uparrow \downarrow\rangle - |\downarrow \uparrow\rangle)\\
&|\psi_{1,+1}\rangle = |\uparrow \uparrow\rangle \\
&|\psi_{1,-1}\rangle = |\downarrow \downarrow\rangle \\
&|\psi_{1,0}\rangle = \frac{1}{\sqrt{2}}(|\uparrow \downarrow\rangle + |\downarrow \uparrow\rangle)
\end{align*}

where the pseudo-spins can be identified as the CEF Kramer's doublet ground states $|\pm\rangle$, i.e., $|\uparrow\rangle = |+\rangle$ and $|\downarrow\rangle = |-\rangle$. These Kramer's doublet wavefunctions can be expressed as linear combinations of $\mathsf{J_z}$ eigenstates, $|J,M_J\rangle$, within a constant $J$ manifold \cite{abragam2012electron}: 
\begin{align*}
 |+\rangle &= \sum_{M_J = -J}^{J} C_{M_J} |J, M_J\rangle, \\
 |-\rangle &= \sum_{M_J = -J}^{J} C^*_{M_J} (-1)^{J-M_J} |J, -M_J\rangle 
\end{align*}
 The neutron scattering intensity for transitions from the ground state to the excited states of an isolated dimer are proportional to:
\begin{equation*}
\begin{split}
I \propto &\langle\psi_{0,0}|\mathsf{J_{z1}}|\psi_{1,\pm1}\rangle^2 + \langle\psi_{0,0}|\mathsf{J_{+1}}|\psi_{1,\pm1}\rangle^2 + \langle\psi_{0,0}|\mathsf{J_{-1}}|\psi_{1,\pm1}\rangle^2 + \ldots \\
&\langle\psi_{0,0}|\mathsf{J_{z2}}|\psi_{1,\pm1}\rangle^2 + \langle\psi_{0,0}|\mathsf{J_{+2}}|\psi_{1,\pm1}\rangle^2 + \langle\psi_{0,0}|\mathsf{J_{-2}}|\psi_{1,\pm1}\rangle^2,
\end{split}
\end{equation*}
where, for example, $\mathsf{J_{z1}}$ indicates the operator acts on site 1. 

As a concrete example, for $|\pm\rangle = |7/2, \pm7/2\rangle$ (which we abbreviate as $|\pm 7/2\rangle$), the first term gives, 
\[
 \frac{1}{2}\bigg(\langle 7/2 | \mathsf{J_{z1}} | \pm 7/2\rangle \langle -7/2 | \pm7/2\rangle - \\ \langle -7/2 | \mathsf{J_{z1}} | \pm 7/2\rangle \langle 7/2 | \pm 7/2\rangle\bigg)^2 = 0,
 \]
 since whenever the left inner product of each term (corresponding to site 1) is non-zero (for instance, in the first term, when working with the upper sign), the right inner product (corresponding to site 2) is zero.
Meanwhile the second term gives,
\[
 \frac{1}{2}\bigg(\langle 7/2 | \mathsf{J_{+1}} | \pm 7/2\rangle \langle -7/2 | \pm7/2\rangle - \langle -7/2 | \mathsf{J_{+1}} | \pm 7/2\rangle \langle 7/2 | \pm 7/2\rangle\bigg)^2 = 0,
 \]
since the raising operator does not connect $|-7/2\rangle$ to $|7/2\rangle$. 
All other terms behave similarly. Meanwhile, by similar arguments, one can see that the intensity for the transition from $\psi_{0,0}$ to $\psi_{1,0}$ is non-zero so long as there is a non-zero overlap of $\langle\pm|\mathsf{J_z}|\pm\rangle$, which is generally expected to be true except in some ``accidental'' cases where $\sum_{M_J}M_J|C_{M_J}|^2$ sums to zero. 

The Kramer's doublet composition for \ybsio\ is currently unknown. Based on the reasoning presented here and our observation of INS intensity only in the $\psi_{1,0}$ mode, we anticipate that the doublet has relatively weak matrix elements for the raising and lowering operators (e.g. $\langle -|\mathsf{J_+}|+\rangle$) compared to $\langle\pm|\mathsf{J_z}|\pm\rangle$. However, the $g$-values inferred based on (nearly) saturated magnetization at $H= 5 $T ($g_a* = 3.2$, $g_b = 2.0$, $g_c = 4.8$) are not strictly Ising-like, implying there are \emph{non-zero} matrix elements $\langle - |\mathsf{J}_\pm|\ +\rangle$.

\section{Specific Heat Fitting}
\label{sec:cp}
Modelling the specific heat arising from the excitation of triplons with dispersion relation $\epsilon(k)$ is a fairly non-trivial problem. In the very low-temperature limit, this can be accomplished~\cite{Quilliam2007GSO} simply by considering a model of non-interaction Bosons, giving
\begin{equation}
C_m(T) = k_B\beta^2 \sum_{\vec{k},\alpha} [\epsilon_\alpha(\vec{k})]^2 
\frac{ e^{\beta\epsilon_\alpha(\vec{k})} }{ [ e^{\beta\epsilon(\vec{k})} -1]^2}.
\end{equation}

As this expression neglects the fact that triplons are hard-core Bosons and interactions must be taken into account, it is not applicable at temperatures approaching the gap energy and higher. To fit the specific heat from low temperature to above the triplon band, one can instead estimate the appropriate specific heat with the following expression,
\begin{equation}
C_\mathrm{m}(T) = \frac{\beta^2}{2} \left[ \frac{ \sum_{\alpha, \vec{k}} [\epsilon_\alpha(\vec{k})]^2 e^{-\beta \epsilon_\alpha(\vec{k})} / \Omega}{1 + \sum_{\alpha, \vec{k}} e^{-\beta\epsilon_{\alpha \vec{k}}}/\Omega} - \left( 
\frac{\sum_{\alpha, \vec{k}} \epsilon_\alpha(\vec{k}) e^{-\beta \epsilon_\alpha(\vec{k})} / \Omega}{1 + \sum_{\alpha, \vec{k}} e^{-\beta\epsilon_\alpha(\vec{k})}/\Omega} \right)^2
 \right]
\end{equation}
which is taken from Refs.~\cite{Gu2000, Troyer1994}. The index $\alpha$ labels the three possible triplon bands and $\Omega$ is the volume of the Brillouin zone. In the case of Heisenberg interactions, the three triplon bands become degenerate in zero field, hence the expression can be simplified to
\begin{equation}
C_\mathrm{m}(T) = \frac{\beta^2}{2} \left[ \frac{ 3\sum_{\vec{k}} [\epsilon(\vec{k})]^2 e^{-\beta \epsilon(\vec{k})} / \Omega}{1 + 3\sum_{\vec{k}} e^{-\beta\epsilon_{\vec{k}}}/\Omega} - \left( 
\frac{3\sum_{\vec{k}} \epsilon(\vec{k}) e^{-\beta \epsilon(\vec{k})} / \Omega}{1 + 3\sum_{\vec{k}} e^{-\beta\epsilon(\vec{k})}/\Omega} \right)^2
 \right]
\end{equation}
In order to fit the data in Fig. 2a of the main document, we use the above expression and the following dispersion relation:
\begin{equation}
\epsilon(\vec{k}) = J_{\text{intra}} + 2J_{\text{inter}}\cos(k_x/2)\cos(k_y/2)
\end{equation}
which fairly effectively reproduces the form of the triplon dispersion measured with inelastic neutron scattering in zero field. The fit of the data is very successful, indicating that fairly isotropic interactions can adequately describe the physics of this system. While slightly anisotropic interactions (which would lift the degeneracy of the triplon bands) could also fit the data, they are not necessary. The resulting exchange constants $J_{\text{intra}} = 0.236$ meV and $J_{\text{inter}} = 0.063$ meV are fairly close to the values obtained from the spin-wave analysis ($J_{\text{intra}}$ = 0.217(3) meV and $J_{\text{inter}}$ = 0.089(1) meV, see section on fitting the field polarized spinwaves below). In any case, we should not expect perfect agreement as this form of the specific heat is an approximation and assumes that the triplon dispersion is independent of temperature, which is unlikely to be true. Whereas the fitting of the specific heat data will primarily be affected by the shape of the dispersion at around 1 K (at the specific heat maximum), the inelastic neutron scattering measurements were performed at much lower temperatures (50 mK).

\section{Ultrasound Velocity Measurements}

\begin{figure}[!h]
\includegraphics[scale = 0.6,clip]{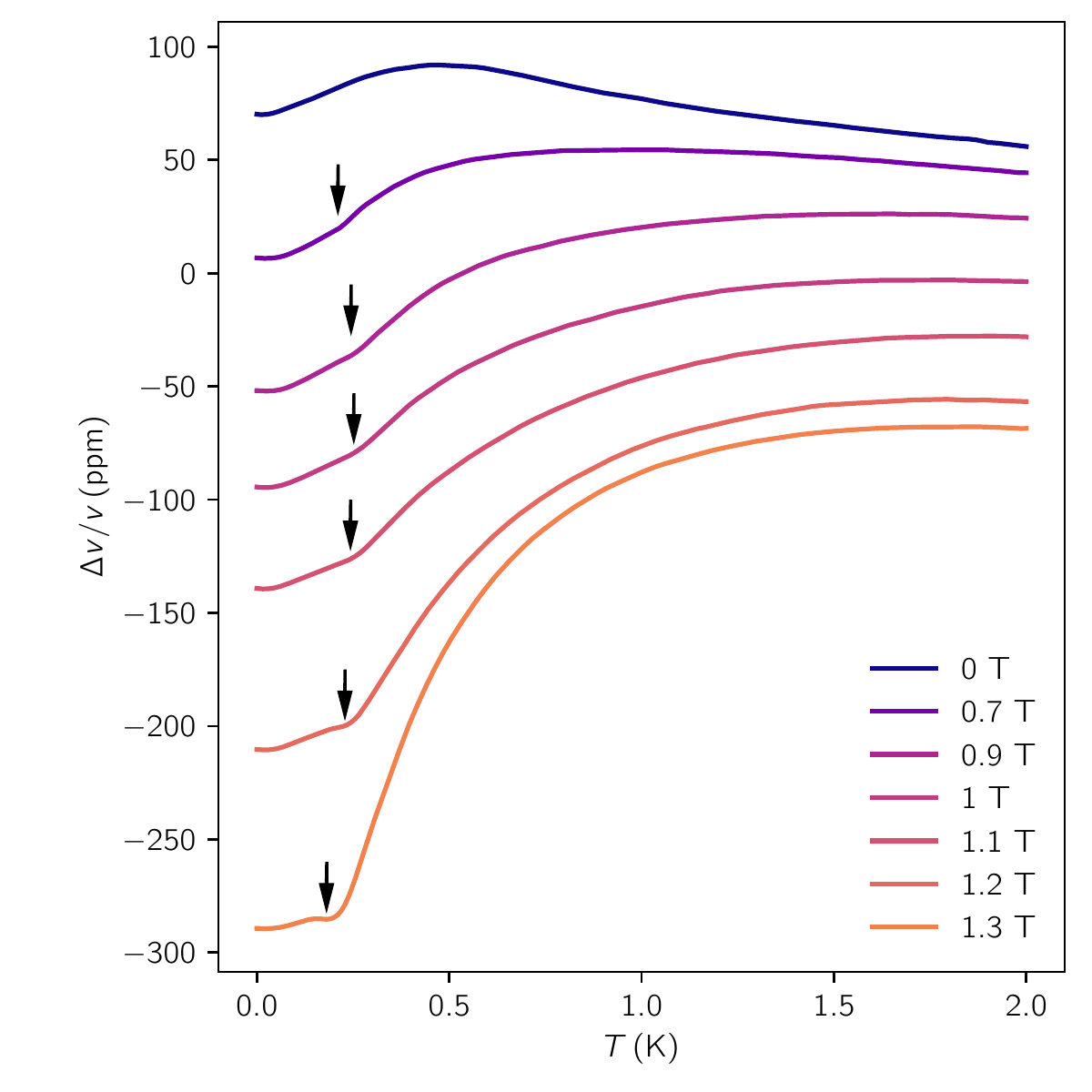}
\caption{Ultrasound velocity measurements as a function of temperature at constant magnetic field. Arrows indicate anomalies associated with antiferromagnetic ordering.} \label{fig_ultrasound}
\end{figure}

Sound velocity measurements were performed as a function of field applied along $c^*$ at fixed temperature (as presented in the main text) and as a function of temperature at fixed field (as presented in Fig. \textcolor{blue}{S}\ref{fig_ultrasound}). The ultrasound velocity experiments were performed down to 50 mK using a pulsed, time-of-flight interferometer. 30 MHz transducers were glued to parallel surfaces so as to propagate longitudinally polarized sound waves along the $c^\ast$-axis. The absolute velocity of the quasi-longitudinal mode studied here was approximately 3000 m/s and relative changes in velocity ($\Delta v/v$) were measured with high precision using a phase-lock loop. Antiferromagnetic phase boundaries could be determined at low temperatures (below 150 mK) from the field sweeps by selecting a sharp change in slope ($H_{c1}$) and a minimum in sound velocity ($H_{c2}$). As the temperature is raised, these anomalies are significantly broadened and it becomes impossible to determine phase boundaries from the field sweeps. The top of the antiferromagnetic ``dome'' was thus determined from small anomalies (abrupt changes in slope) in the temperature sweeps shown in Fig. \textcolor{blue}{S}\ref{fig_ultrasound}. These anomalies are entirely consistent with the peaks found in low-temperature specific heat measurements.

The inset of Fig. 4c in the main text shows a comparison of the sound velocity field sweep with the field-derivative of the neutron Bragg intensity, $dI/dB$, which is proportional to $d(m_z^2)/dB = 2m_z\chi$. The agreement is excellent and this suggests that the sound velocity is well coupled to the uniform longitudinal magnetization. However, a standard theoretical treatment gives a somewhat different relationship between magnetization and sound velocity. Assuming a linear-quadratic magnetoelastic coupling term in the free energy
\[ F_{me} = \frac{1}{2}\kappa \epsilon_n m_z^2 \]
where $\kappa$ is a magnetoelastic coupling constant related to a particular element of the strain tensor, $\epsilon_n$. Following the work of Quirion \emph{et al.}~\cite{Quirion2011}, amongst others, the elastic constant is renormalized as
\[ C_{mn} = \frac{\partial^2 F}{\partial\epsilon_m\partial\epsilon_n} - \frac{\partial^2 F}{\partial m_z\partial \epsilon_m}
\left( \frac{\partial^2 F}{\partial m_z^2}\right)^{-1} \frac{\partial^2 F}{\partial m_z\partial \epsilon_n} \]
For the particular mode studied here,
\[ C_{33} = C_{33}^0 - \left( \frac{\partial^2 F}{\partial m_z\partial \epsilon_3}\right)^2
\left( \frac{\partial^2 F}{\partial m_z^2}\right)^{-1} \]
\[ C_{33} = C_{33}^0 - (2\kappa m_z)^2 a^{-1} = C_{33}^0 - \kappa^2m_z^2\chi_z \] 
Relative changes in sound velocity can then be related to relative changes in elastic constant through
\[ \frac{\Delta v}{v} = \frac{ \Delta C_{33}}{2C_{33}^0} = - \frac{\kappa^2m_z^2\chi}{2C_{33}^0} \]

There are thus two striking differences between this simple theory and the results. 1. As mentioned above, experimentally 
$\Delta v/v \propto |m|\chi$ whereas the theory predicts $\Delta v/v \propto m^2\chi$. As such, the dip at $H_{c2}$ is less pronounced experimentally than theoretically. 2. The area under the curve $\int_0^{B_\mathrm{sat.}} (\Delta v/v)dB$, which according to theory should simply give $m_\mathrm{sat.}^3$, is in reality strongly temperature dependent. Hence, in the future, a more elaborate theoretical treatment of sound velocity for such a system, including coupling to the antiferromagnetic order parameter and spin fluctuations, would be valuable and might provide a more quantitative understanding of these results.

\section{Single Crystal Neutron Scattering}
\begin{figure}[!h]
\includegraphics[scale = 0.8,clip]{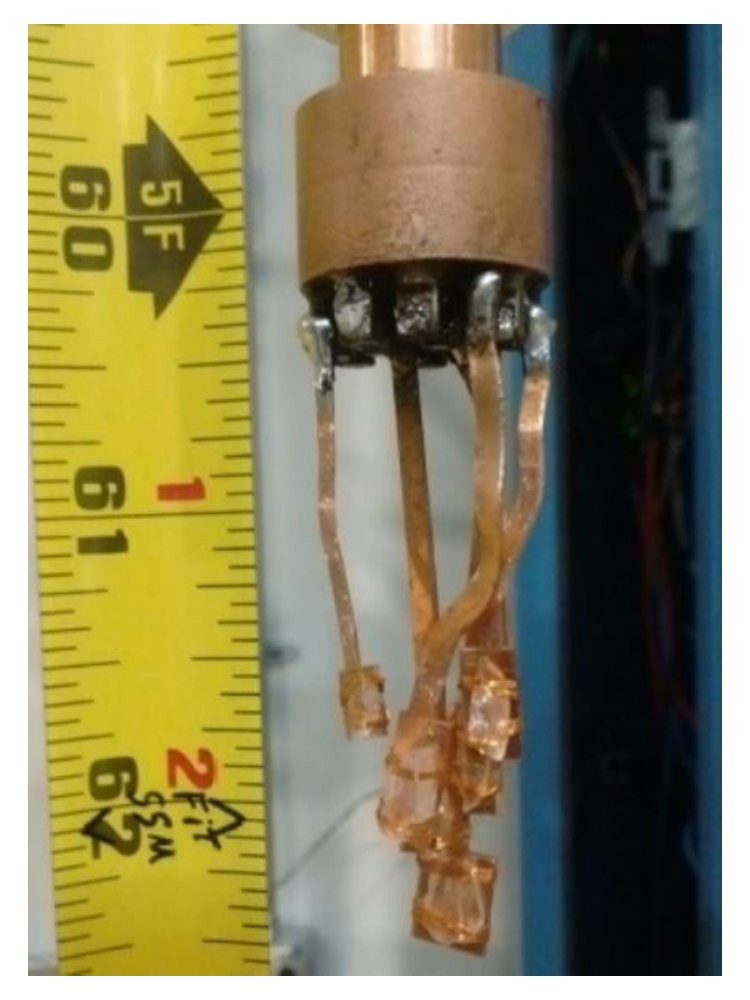}
\caption{The coalignment of five crystals used for the inelastic neutron scattering measurement.} \label{fig_mount}
\end{figure}

\begin{figure*}[!htb]
\includegraphics[scale = 0.45,clip]{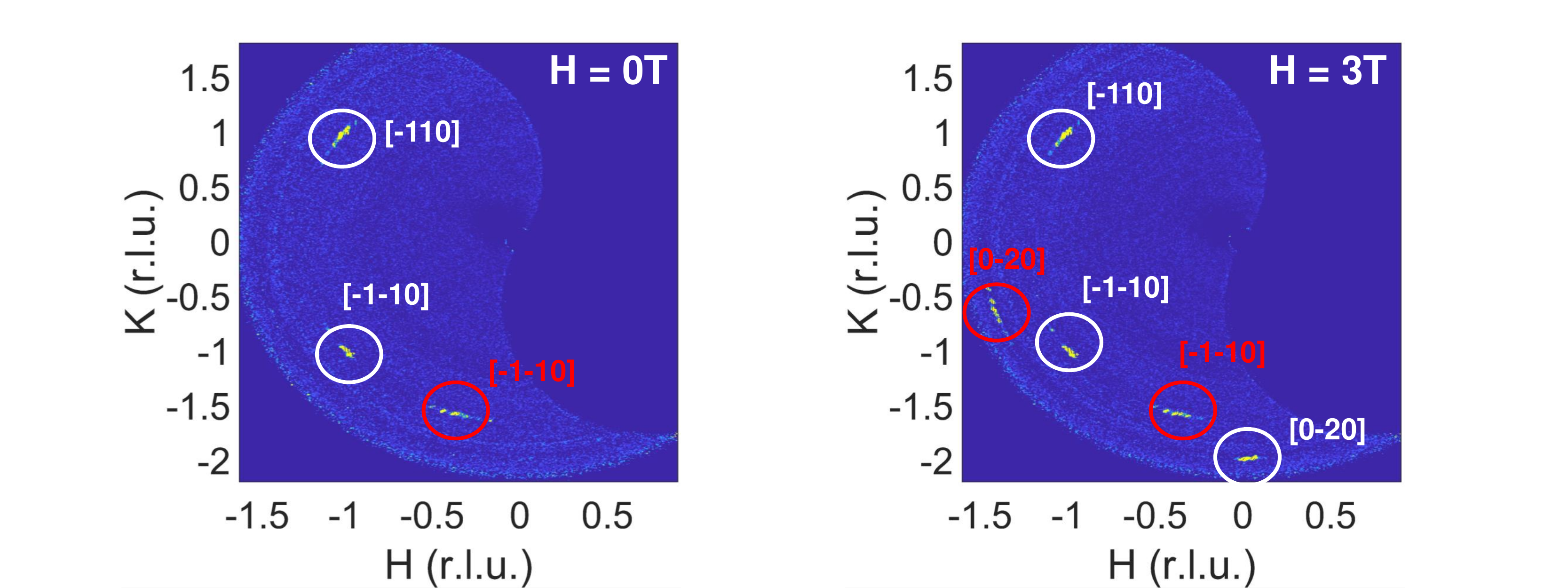}
\centering
\caption{Slices taken at the elastic line (integrated from $E=-0.1$ to 0.1 meV) with $E_i$ = 1.55 meV. Bragg peaks (both nuclear and magnetic in origin) arising from the main grain are labeled in white. Bragg peaks from the misaligned grain are circled in red. The HKL values for reflections from the misaligned grain were determined based on their 2$\theta$ values.} \label{fig_Elastic}
\end{figure*}

\begin{figure}[!h]
\includegraphics[scale = 0.25]{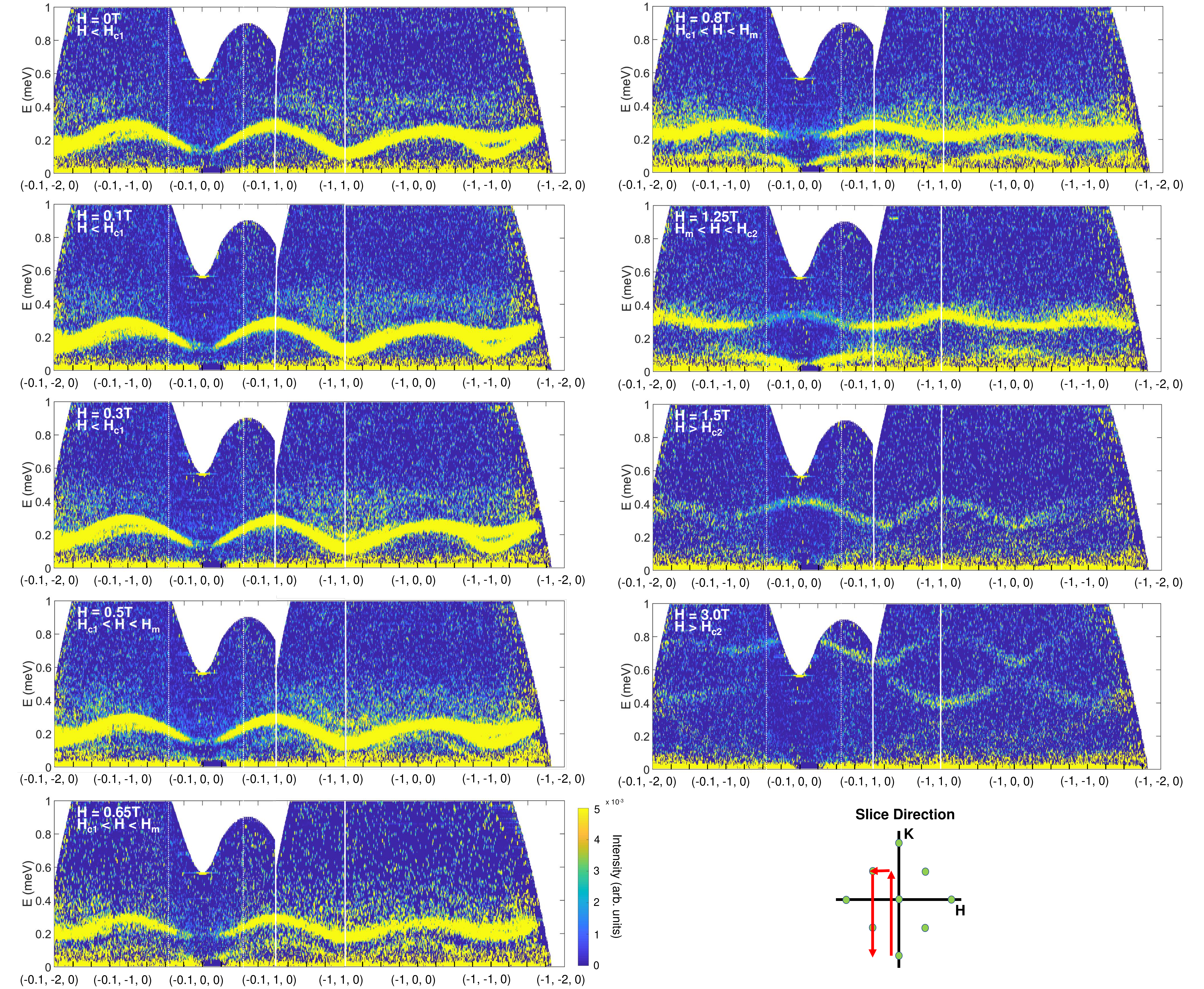}
\noindent
\centering
\caption{INS data at 50mK shown for an "open-rectangular" path of reciprocal space (shown in the bottom right of the figure), for all the magnetic field strengths measured. } \label{fig_spaghettiplot}
\end{figure}

Five single crystals were co-aligned using Laue X-ray scattering to achieve an overall mass of 1.1 g and a mosaic spread of less than 3$^{\circ}$ of the dominant grain. The crystal mount is shown in Fig. \textcolor{blue}{S}\ref{fig_mount}. One of the crystals was later discovered to contain a misaligned grain, which is visible in the neutron scattering data. The elastic scattering (-0.1 meV to 0.1 meV) shows Bragg peaks from the misaligned grain, highlighted by red circles in Fig. \textcolor{blue}{S}\ref{fig_Elastic}. Figure \textcolor{blue}{S}\ref{fig_spaghettiplot} shows inelastic slices for every applied magnetic field setting ($H||c$). The path through reciprocal space shown in these plots is illustrated to the right of the figure. At low fields, the presence of the misaligned grain is clearly observed, manifesting as what looks like an additional excitation branch in portions of the HK0 plane. It is particularly prevalent at ($\overline{0.1}$,$\overline{1}$,0) and ($\overline{1}$,$\overline{1}$,0). The excitations of this misaligned grain are not visible for fields greater than 1.25 T. This may be due to the overall decrease in inelastic intensity which occurs due to the development of strong magnetic Bragg peaks. Additionally, for field values near \hcone\ the $S_{z} = \pm 1$ modes (which we have called $\psi_{1,\pm1}$ above) are (barely) visible near the ($\overline{1}$,$1$,0) reciprocal lattice point (also shown in Fig. S6 as line cuts). This indicates that the aforementioned transition matrix elements from the ground state to the $\psi_{1,\pm1}$ states are small but non-zero.

\newpage

\clearpage

\begin{figure}[!htb]
\includegraphics[scale = 1]{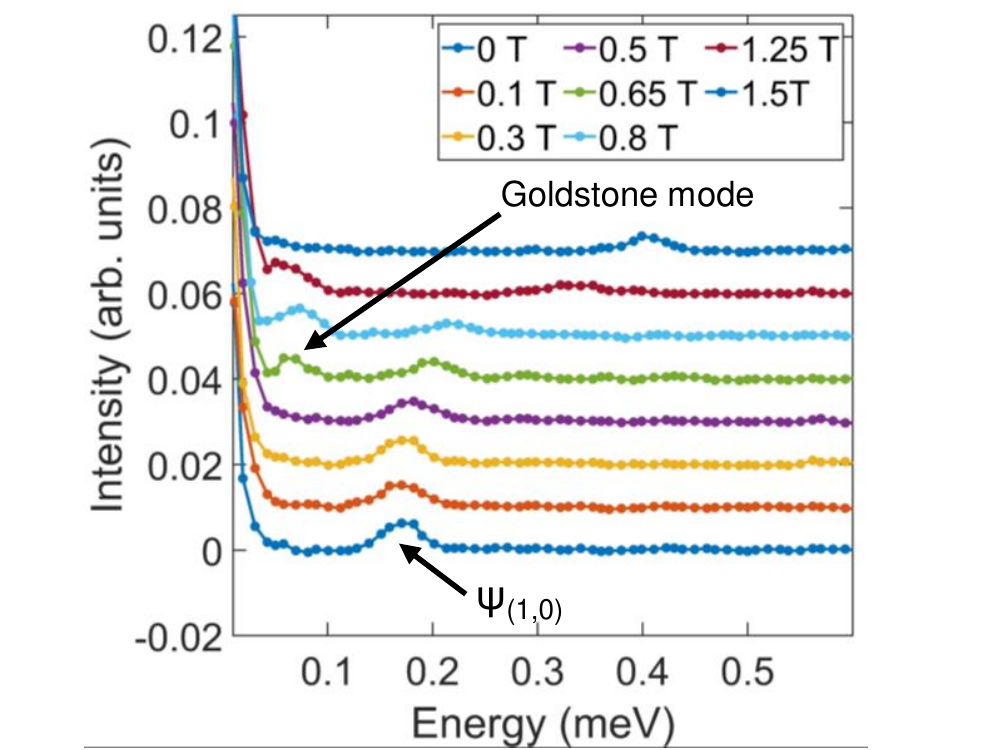}
\noindent
\centering
\caption{Intensity vs. energy cuts of the inelastic neutron spectrum at $|Q|$ = $0.2362$ \AA$^{-1}$. The excitation branch \emph{leading} to the Goldstone mode is visible as a low energy peak between \hcone\ (~0.4 T) and \hctwo\ (~1.4 T). Below \hcone, the peak near 0.19 meV remains at constant energy, identifying it as a S$_z$ = 0 excitation ($\psi_0$)} \label{fig_GMcut}
\end{figure}

Fig. S\ref{fig_GMcut} shows line cuts taken at $|Q|$ = $0.2362$ \AA$^{-1}$ for all the measured field strengths. The primary excitation ($\psi_{1,0}$) does not change in energy below \hcone\ indicating its angular momentum projection along the magnetic field is 0. Thus we conclude that this excitation is the excitation to the S$_{tot}$ = 0, S$_z$ = 0 mode. Additionally, the development of the branch connected to the Goldstone mode is visible at $\sim$ 0.08 meV and is only observed for fields between \hcone\ and \hctwo. 

\begin{figure}[!htb]
\includegraphics[scale = 1]{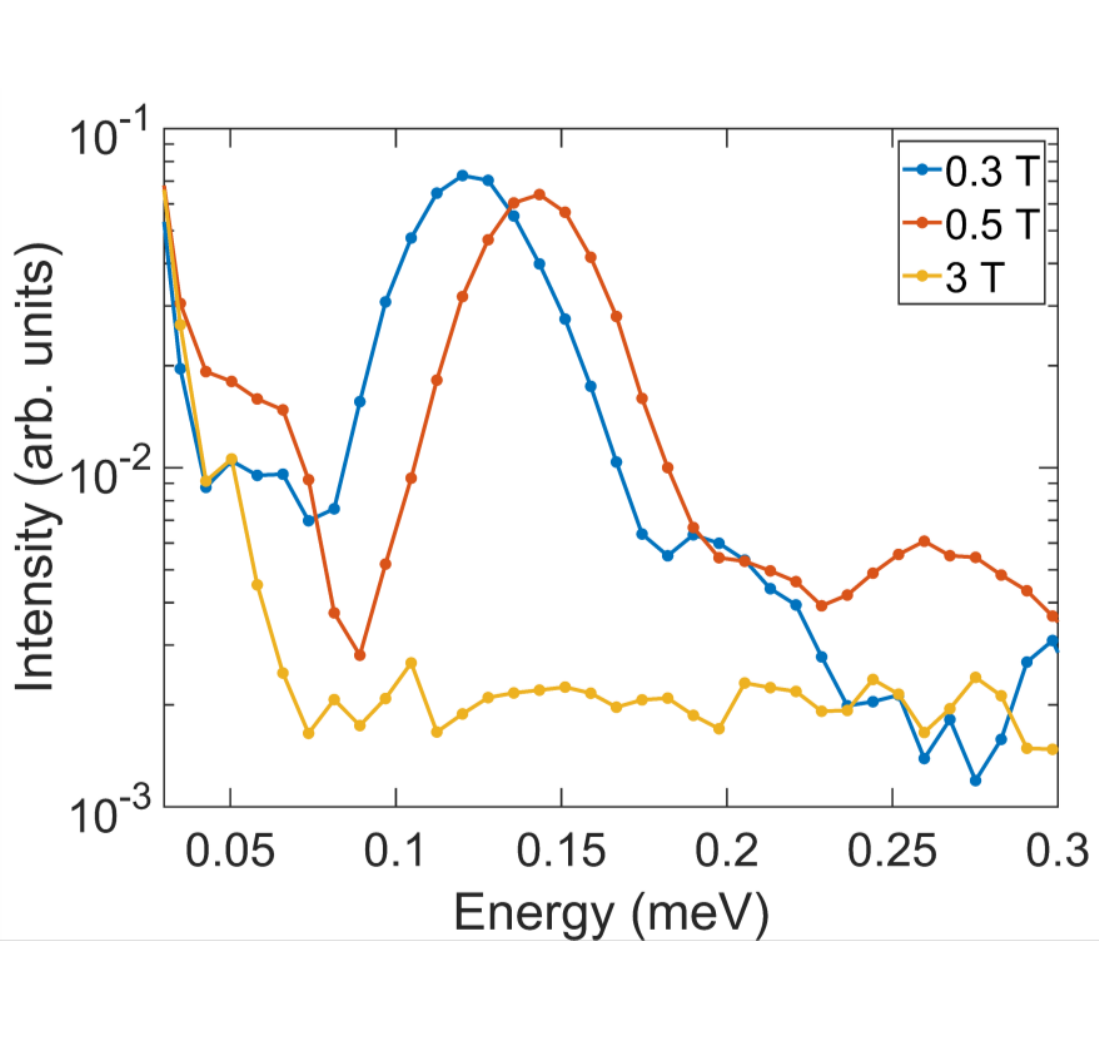}
\noindent
\centering
\caption{Cuts of the inelastic neutron spectrum at (-1, 1, 0), integrating over H = [-1.1,-0.9] and K = [0.75,1.25] r.l.u. for three different magnetic field strengths. The sharp large peaks in 0.3 T and 0.5 T are the S$_z$ = 0 mode observed at all field values. On the lower and higher energy sides of this peak additional features from the S$_z$ = $\pm 1$ modes are observed.} \label{fig_Triplet}
\end{figure}

Fig. S\ref{fig_Triplet} shows line cuts on a logarithmic intensity scale, taken at ($\overline{1}, 1$, 0) for three different magnetic field strengths: 0.3 T, 0.5 T, and 3 T. The 3 T data is shown as a reference to what the expected background would be for this energy range. At 0.3 T and 0.5 T the main $\psi_{1,0}$ excitation is observed at $\sim$ 0.12 meV and $\sim$ 0.14 meV, respectively. At H = 0.3T and 0.5T, two weak excitations are seen to split off of the main line, which can likely be identified as the $\psi_{1,\pm 1}$ states. The weak intensity of these modes is likely due to the effect of the crystal electric field on the matrix elements for the transitions, as discussed above. 

\section{Fits to Field-Polarized Spin Waves}
The spin wave dispersions measured in the field-polarized limit ($H = 3$T) were fit using linear spin wave theory (LSWT) as implemented by the SpinW package \cite{Toth2015}, which evaluates the goodness of fit based on agreement between the measured and calculated dispersions (and does not include comparisons of intensities). Several types of fits were attempted, as described further below. Each fit was performed using the "particle swarm optimizer" algorithm for 10 runs with 100 iterations per run, and a maximum number of function evaluations of 1$\times 10 ^5$. The outputs for each fit are shown in Table \ref{tab_fits}.

\begin{table}[!h]
 \renewcommand*{\arraystretch}{2}
 \begin{tabular}{| c | c | c | c | c |} 
 \hline
 & Unconstrained Heisenberg & Constrained Heisenberg & XXZ/Heisenberg & Specific Heat \\ 
 \hline
 \Jra\ (meV) & 0.18(2) & 0.217(3) & $J_{\text{intra, XX}}$ = 0.190(3), $J_{\text{intra, Z}}$ = 0.180(4) & 0.236(4) \\ 
 \hline
 \Jer\ (meV) & 0.12(1) & 0.089(1) & 0.121(1) & 0.06(2)\\
 \hline
 $g_{zz}$ & 4.82(5) & 4.68(1) & 4.8 (fixed) & N/A\\
 \hline
\end{tabular}
 \caption{Parameters from the various types of fits (and comparing to the result of fitting the zero field $C_p$). The quoted error for the parameters extracted from spin wave fits is from the standard deviation of the respective parameters in the 10 runs. \label{tab_fits}}
\end{table}

The magnetic structure was optimized for each new set of trial parameters (our addition to the pre-existing SpinW fitting routine). The reference frame chosen by SpinW is $x = a$, $y = b$, and $z=c^*$, so for our fits involving an XXZ form of the interaction Hamiltonian (where we have assume $z=c$, i.e. the field direction, which we call the ``experimental coordinate frame''), we performed a coordinate transformation on the XXZ exchange matrix and constrained the variation of the resulting (non-diagonal) matrix elements to ensure XXZ symmetry in the experimental frame. \textbf{All of our exchange parameters and symmetries mentioned below and in the main text are with respect to the $a^*$, $b$, $c$ basis, i.e., the experimental frame.} Fits were performed with data taken from the following slices: (-1K0), (H00), (H10), (-HH0), (H-2H0), and (-0.1K0). The spin waves were fit to the Hamiltonian below:

\[H = \sum_{<i,j>}\vec{S_{i}}\ \bar{J}_{\text{intra}} \vec{S_{j}} + \sum_{<<i,j>>}\vec{S_{i}}\ \bar{J}_{\text{inter}} \vec{S_{j}} + \sum_{i}\vec{B}\ \bar{g}\ \vec{S_{i}}
\]

where \Jintra\ and \Jinter\ are the intra- and interdimer exchange interaction tensors as labeled in Fig. 1a of the main text, and $\bar{g}$ denotes the $g$-tensor. For the two exchange tensors, the lowest possible symmetry is (based on the space group symmetries):

\[\bar{J}_{\text{intra}} = 
 \left[ {\begin{array}{ccc}
 A_1 & 0 & D_1 \\
 0 & B_1 & 0 \\
 D_1 & 0 & C_1 \\
 \end{array} } \right]
\]

\[\bar{J}_{\text{inter}} = 
 \left[ {\begin{array}{ccc}
 A _2& E & D_2 \\
 E & B_2 & F \\
 D_2 & F & C_2 \\
 \end{array} } \right]
\]

\begin{figure}[!h]
\includegraphics[scale = 0.21]{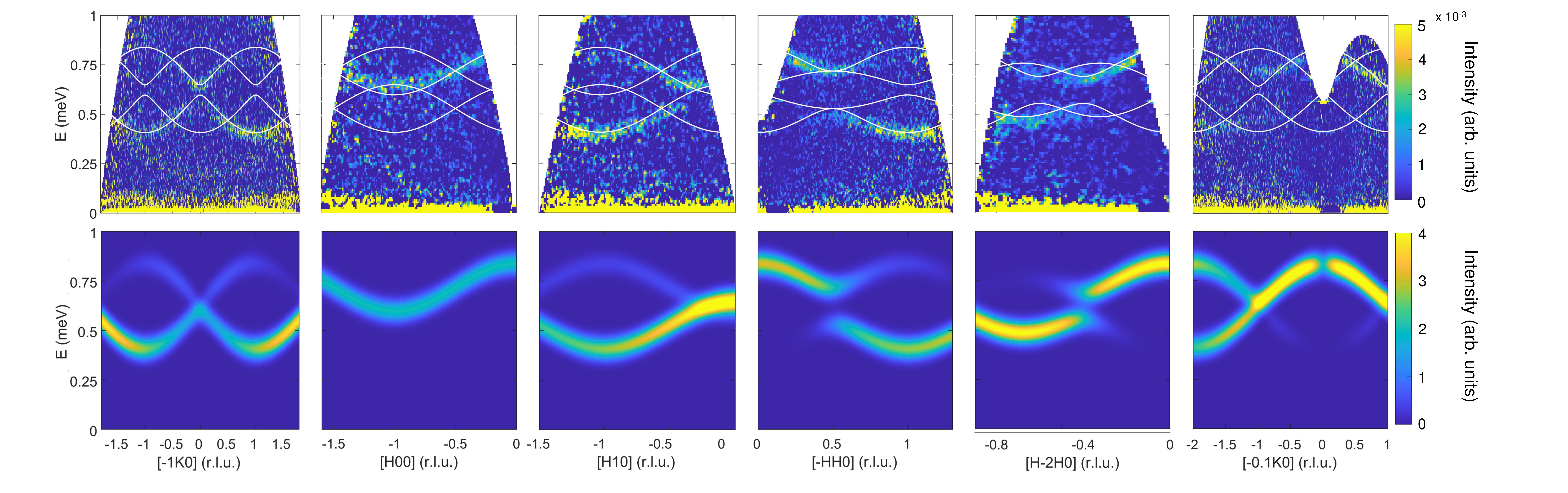}
\centering
\caption{(top row) Overlays of dispersion on the inelastic neutron scattering data using the \Jra\ XXZ and \Jer\ Heisenberg fit parameters. The extracted parameters from this fit are: $J_{\text{intra, XX}}$ = $0.190(3)$ meV, $J_{\text{intra, Z}}$ = $0.180(4)$ meV, and \Jer\ = $0.121(1)$ meV. There are four bands due to having 4 magnetic atoms in the unit cell of \ybsio. (bottom row) Calculated neutron spectra for the same set of parameters.} \label{fig_XXZandHeisen}
\end{figure}

Using these full symmetry-allowed forms would constitute fitting 10 independent parameters, which for our data set is infeasible. Therefore, we started with the lowest possible symmetry that one would expect given the observation of a Goldstone mode with the field applied along the $c$-axis, which is an XXZ type interaction. However, we did not fit \Jinter\ as XXZ due to a direct correlation between $J_{\text{intra, Z}}$ and $J_{\text{inter, Z}}$. With this in consideration, we fit the exchange interactions as XXZ for \Jra\ and Heisenberg for \Jer\ shown in Fig. S\ref{fig_XXZandHeisen}. The extracted parameters are: $J_{\text{intra, XX}}$ = $0.190(3)$ meV, $J_{\text{intra, Z}}$ = $0.180(4)$ meV, and \Jer\ = $0.121(1)$ meV. Additionally, we would like to remind the reader that we are using the convention of positive $J$ meaning antiferromagnetic exchange. This fit shows good qualitative agreement, however, along certain directions (such as (H00), (H10), and (-0.1K0)) it fails to reproduce some intensity features. Additionally, the shape of the dispersions do not exactly match, which is particularly evident along (H00) and (-0.1,K,0). The reason for this disagreement is uncertain, but suggests that weaker in-plane anisotropies are responsible.

\begin{figure}[!h]
\includegraphics[scale = 0.21]{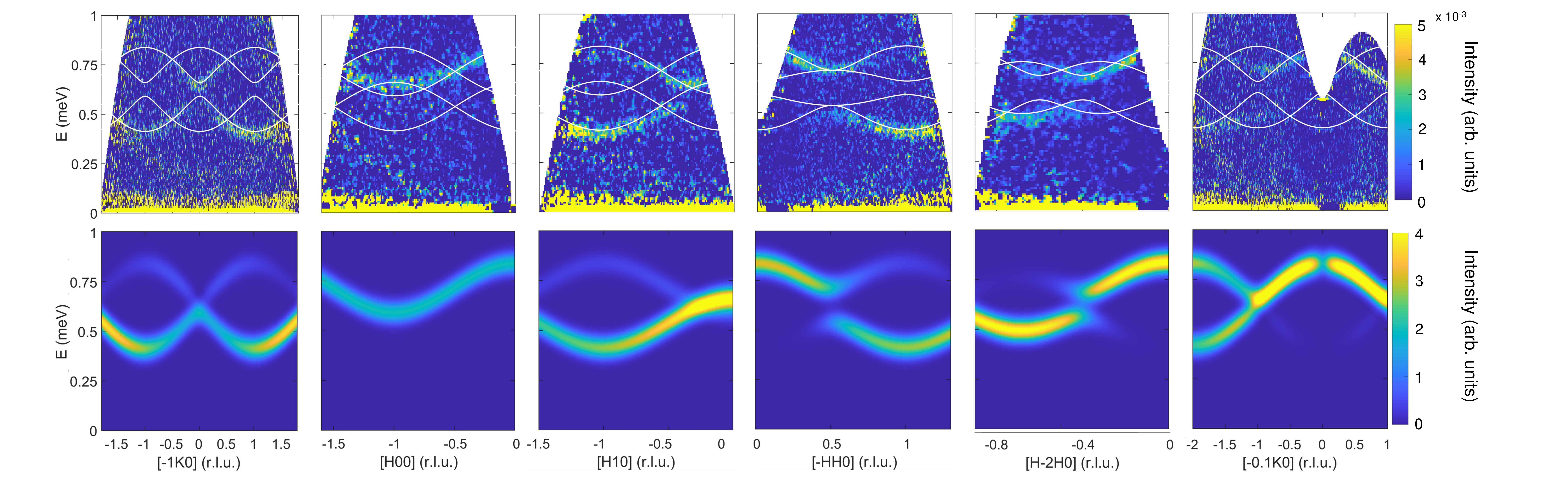}
\centering
\caption{(top row) Overlays of dispersion on the inelastic neutron scattering data using the \Jra\ Heisenberg and \Jer\ Heisenberg fit parameters. The extracted parameters from this fit are: \Jra\ = $0.18(2)$ meV, \Jer\ = $0.12(1)$ meV, and $g_c$ = $4.82(5)$. There are four bands due to having 4 magnetic atoms in the unit cell of \ybsio. (bottom row) Calculated neutron spectra for the same set of parameters.} \label{fig_Unconstrained}
\end{figure}

We also include fits for Heisenberg symmetry on both \Jra\ and \Jer. In these fits we also let the z component of the g-tensor $g_c$ vary $\pm 0.5$ from the value we infer from magnetization measurements (4.8). The extracted parameters from the fit shown in Fig. S\ref{fig_Unconstrained} are: \Jra\ = $0.18(2)$ meV, \Jer\ = $0.12(1)$ meV, and $g_c$ = $4.82(5)$. Visually, the Heisenberg fit produces similar results to the XXZ fit, which is as expected since the intensities and dispersions of the in-plane (HK0) spin waves should be determined by the in-plane interactions only (the z exchange acts like $g_c$ and serves only to shift the bands up and down).

Considering qualitative agreement with the measured intensities (rather than just the dispersion relations), we found that Heisenberg parameters constrained in the following way could provide a better agreement: \Jra\ constrained between $0.18$ and $0.3$ while \Jer\ was constrained between $0$ and $0.1$. This range was roughly determined by manually adjusting the parameters and observing how the intensities of the spectra changed. The parameters from this constrained fit are: \Jra\ = $0.217(3)$ meV, \Jer\ = $0.089(1)$ meV, and $g_c$ = $4.68(1)$. This was the most consistent fit we obtained considering both dispersions and intensities. These constrained fits are shown in Fig. S\ref{fig_Constrained} below. While the same issues exist regarding the agreement of dispersion relations for (H,0,0) and (-0.1,K,0) directions, the intensity of trade-off between the upper and lower branches along (-1,K,0) is captured better.  In the main text we have adopted these parameters obtained from this version of the fit.

\begin{figure}[!h]
\includegraphics[scale = 0.21]{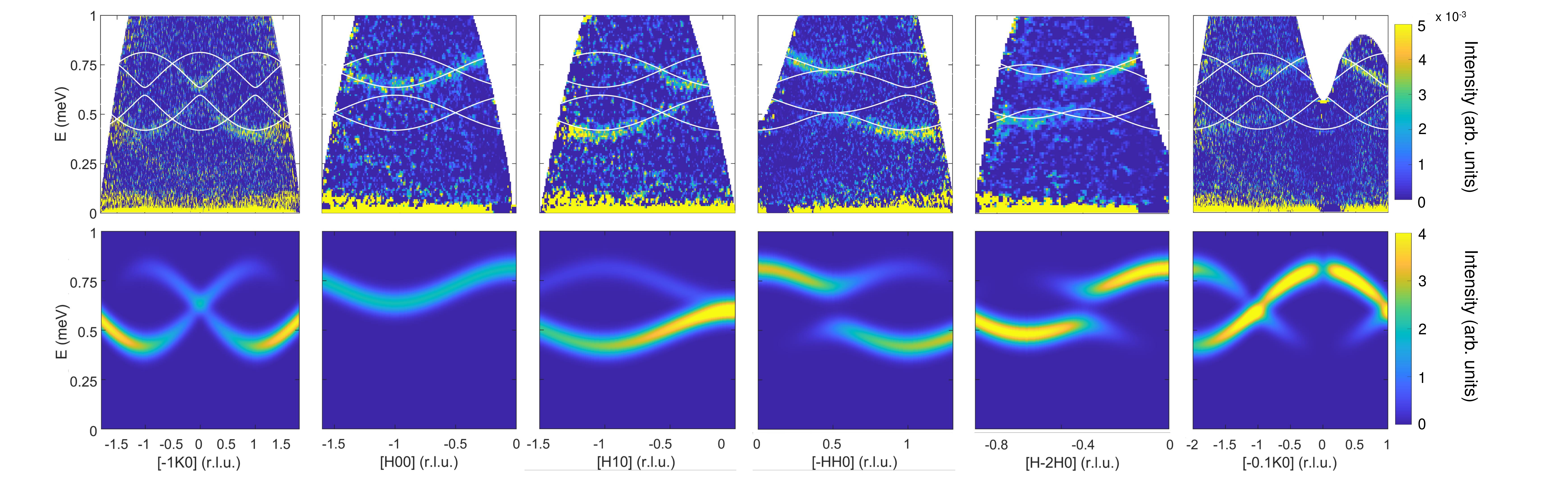}
\centering
\caption{(top row) Overlays of dispersion on the inelastic neutron scattering data using the \Jra\ Heisenberg and \Jer\ Heisenberg fit parameters that were constrained in fitting. The parameters from this fit are: \Jra\ = $0.217(3)$ meV, \Jer\ = $0.089(1)$ meV, and $g_c$ = $4.68(1)$. There are four bands due to having 4 magnetic atoms in the unit cell of \ybsio. (bottom row) Calculated neutron spectra for the same set of parameters.} \label{fig_Constrained}
\end{figure}

\begin{figure}[!htb]
\includegraphics[scale = 0.9]{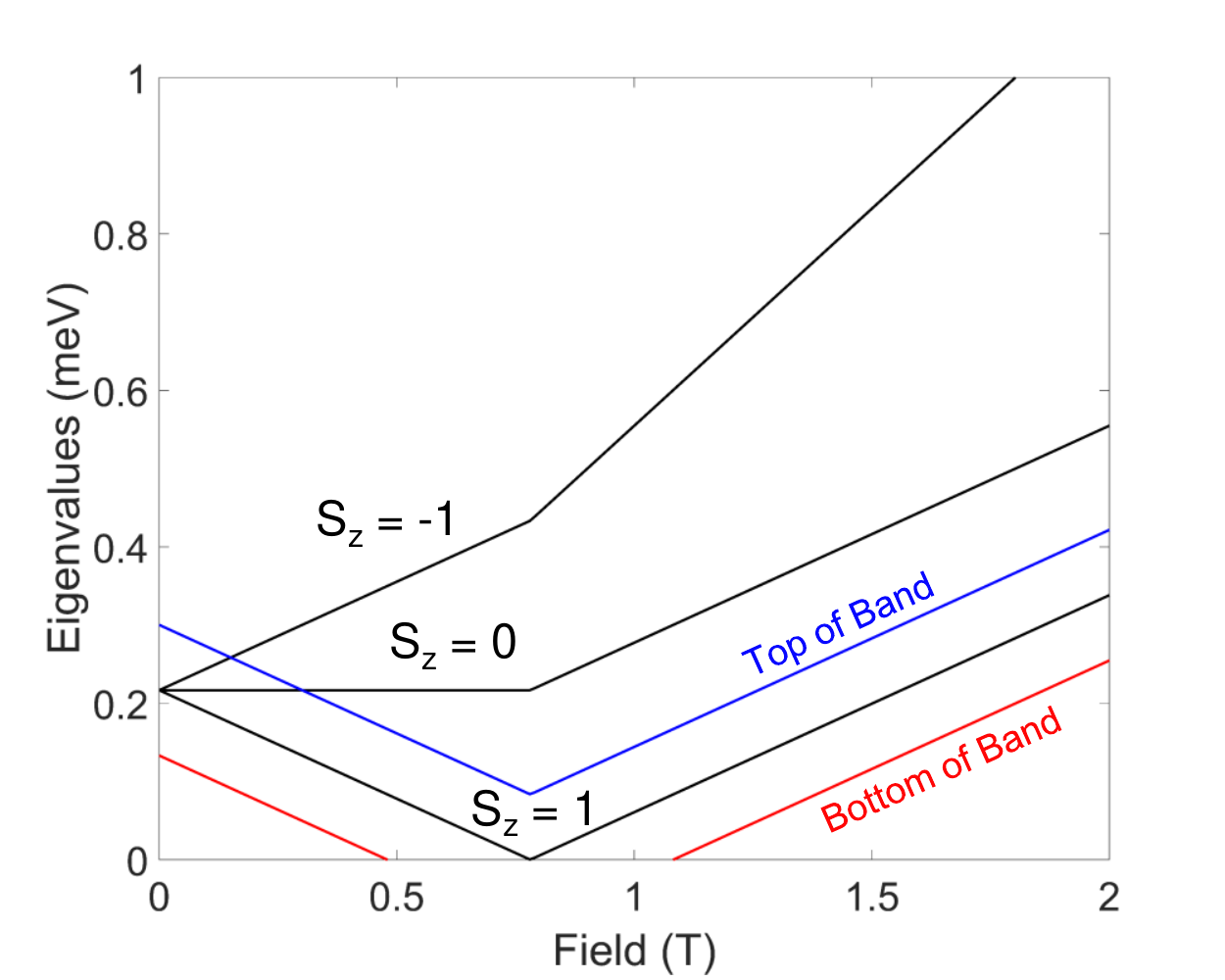}
\centering
\caption{Calculation of the eigenvalues of the triplet modes with an isolated dimer model using the parameters from the constrained Heisenberg fit (Fig. S\ref{fig_Constrained}). The upper and lower bound of the S$_Z$ = 1 mode are shown using the bandwidth determined experimentally, as in this model they determine \hcone\ and \hctwo.} \label{fig_DimerSplitting}
\end{figure}

In addition to the aforementioned improvements, the fit from Fig. S\ref{fig_Constrained} also provides a more realistic \hcone\ value if an isolated dimer model is considered. The calculation of the isolated dimer model triplet splitting is shown in Fig. S\ref{fig_DimerSplitting}, where the lower band and upper band lines are determined from the measured dispersion of the H = 0 T spin wave excitation (see main text). 

\section{Powder Neutron Diffraction}
\begin{figure}[!htb]
\includegraphics[scale = 0.6]{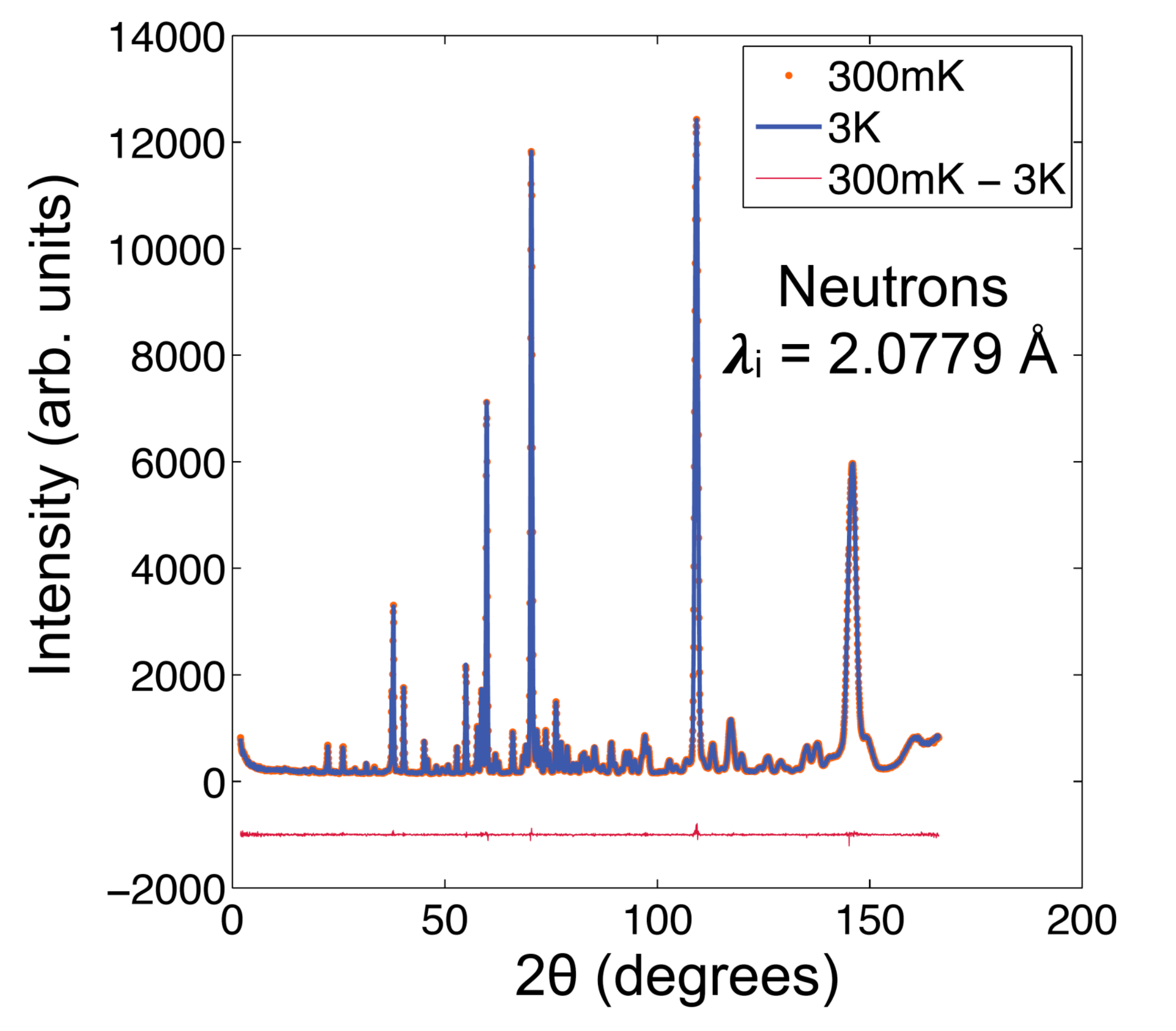}
\centering
\caption{Powder neutron diffraction data taken at 300 mK and 3 K. No additional Bragg peaks are observed at 300 mK confirming the non-magnetic singlet ground state.} \label{fig_BT1}
\end{figure}

Neutron powder diffraction data (taken on the instrument BT1 at the NIST Center for Neutron Research) confirms a lack of long range magnetic order in zero applied field. No magnetic Bragg peaks are observed at 300 mK (see the high vs. low temperature difference plot in Fig. \textcolor{blue}{S}\ref{fig_BT1}).
\section{Powder Synchotron X-Ray Diffraction}
Synchrotron X-ray powder diffraction was performed using the Advanced Photon Source at Argonne National Laboratory using the 11-BM beamline with $\lambda$ = 0.4122\AA. Rietveld refinement of the data agrees well with the previously published structure of \ybsio\ \cite{Felsche1970}. Refined structural parameters at 295 K are listed in Table \ref{tab:params}. 

\begin{figure}[h!]
\includegraphics[scale = 0.4]{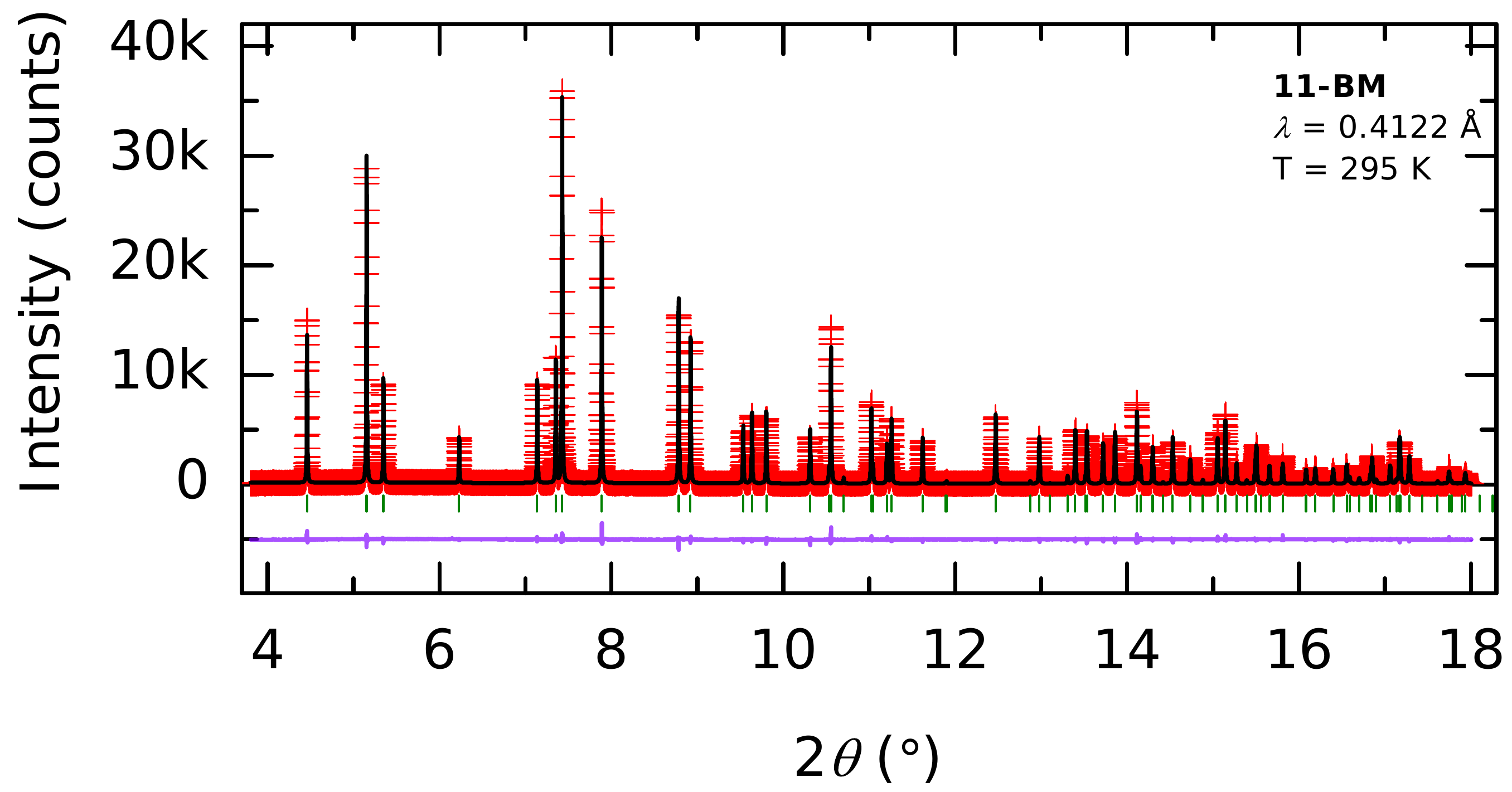}
\centering
\caption{Synchrotron X-ray diffraction data measured on polycrystalline \ybsio\ at 295 K (red data points), with structural refinement (black line) and difference curve (purple line) shown.} \label{fig_11BM}
\end{figure}

\begin{table}[!h]
 \begin{tabular}{||c c c c c||} 
 \hline
 Atom & Site & x & y & z \\ [0.5ex] 
 \hline\hline
 Yb & 4$g$ & 0.5 & 0.8066(8) & 0 \\ 
 \hline
 Si & 4$i$ & 0.7184(4) & 0.5 & 0.4137(6) \\
 \hline
 O & 2$c$ & 0.5 & 0.5 & 0.5 \\
 \hline
 O & 4$i$ & 0.8804(7) & 0.5 & 0.7215(9) \\
 \hline
 O & 8$j$& 0.7362(8) & 0.6510(8) & 0.2160(4) \\ [1ex] 
 \hline
\end{tabular}
 \caption{Structural parameters at 295 K. Space group C2/m, $a$ = 6.7714(9), $b$ = 8.8394(2), $c$ = 4.6896(5), $\beta$ = 101.984(9). \label{tab:params}}
\end{table}



\section{Magnetization}
\begin{figure}[!htb]
\includegraphics[scale = 0.8]{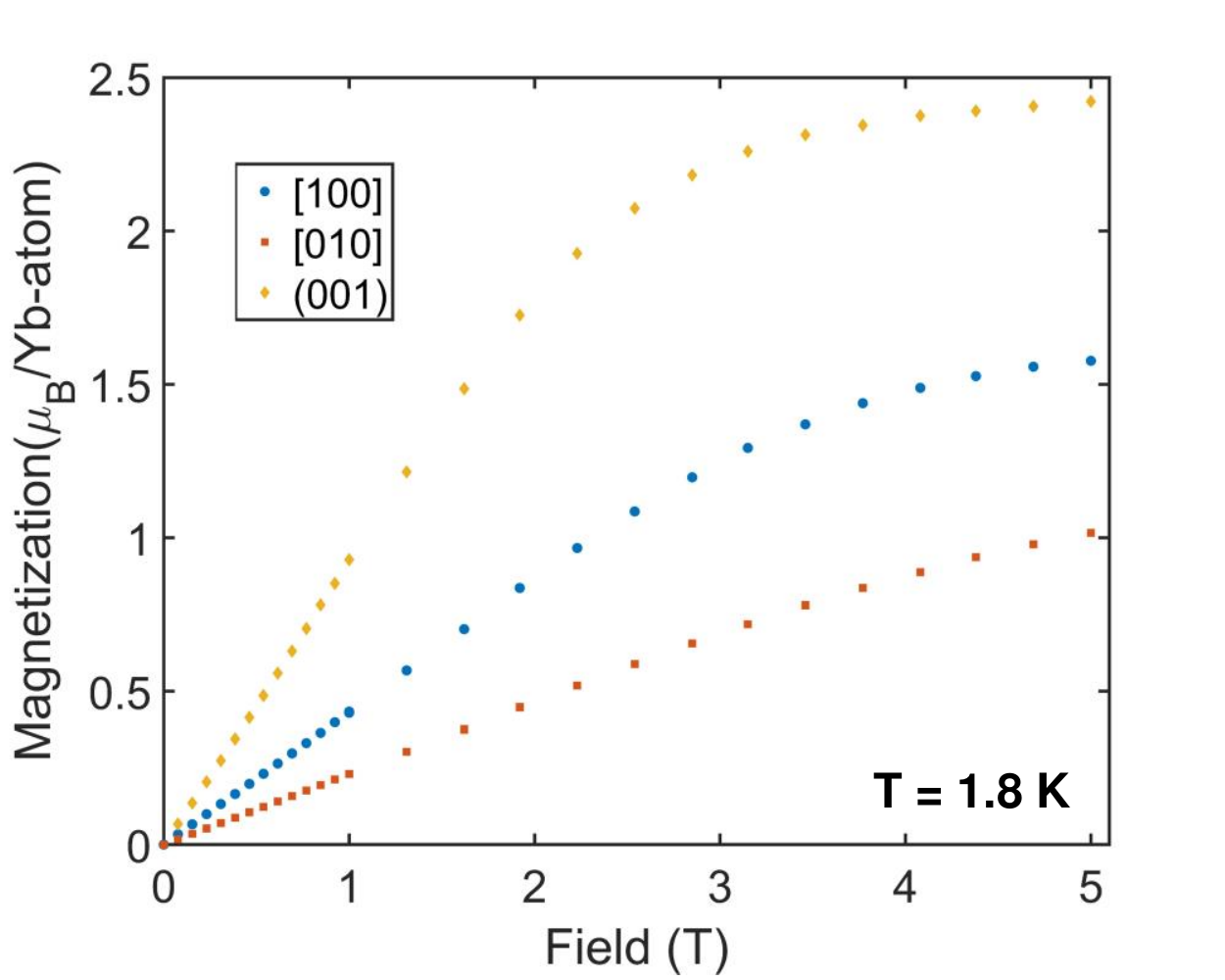}
\centering
\caption{Magnetization performed at 1.8 K along $a^*$, $b^*$, and $c$ yields the $g$-tensor values shown in the main text, $g_{a*}~=~3.2$, $g_{b*}~=~2.0,$ and $g_{c}~=~4.8$. } \label{fig_magnetometry}
\end{figure}

Magnetization was measured at 1.8 K using a Quantum Design SQUID magnetometer, shown in Fig. \textcolor{blue}{S}\ref{fig_magnetometry}. Proper orientation of the crystal was confirmed before and after measurements using Laue x-ray diffraction.
\section{High-Temperature Specific Heat}
\begin{figure}[!htb]
\includegraphics[scale = 0.45]{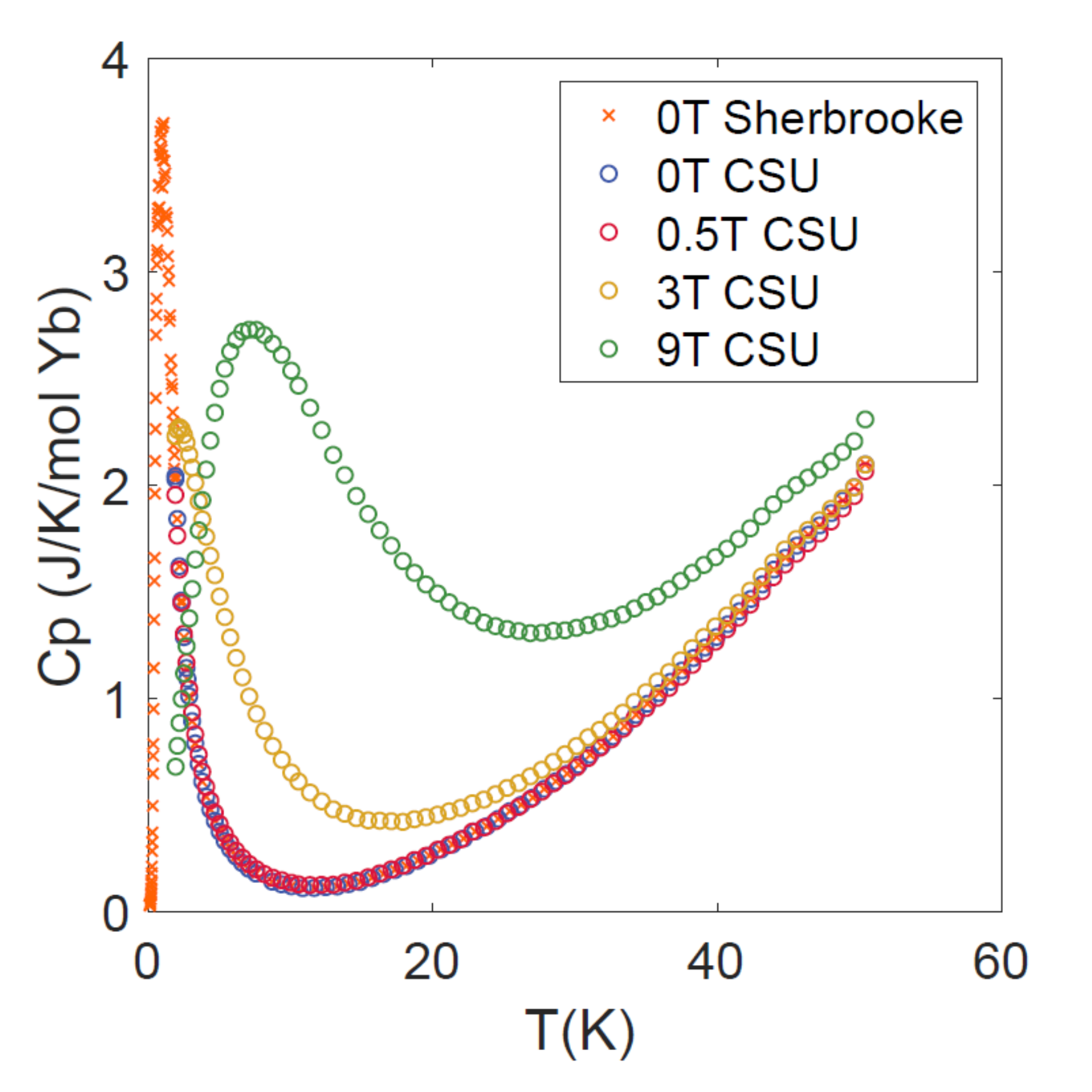}
\centering
\caption{Field-dependent magnetic specific heat (lattice subtracted) from a polycrystalline sample.} \label{fig_Cp}
\end{figure}

The magnetic specific heat of a polycrystalline sample of \ybsio\ was found by subtracting the specific heat of Lu$_2$Si$_2$O$_7$ (the non-magnetic lattice analog), and is shown in Fig. \textcolor{blue}{S}\ref{fig_Cp}. The data shows the field dependence of the low-temperature Schottky anomaly, and also show the beginnings of a separate Schottky anomaly (upturn after 10 K) signaling the presence of a crystal field level at approximately 120 K. The 0T data was reproduced at both Colorado State University (using a Quantum Design Physical Properties Measurement System) and Universit\'e de Sherbrooke.

\begin{figure}[!htb]
\includegraphics[scale = 0.45]{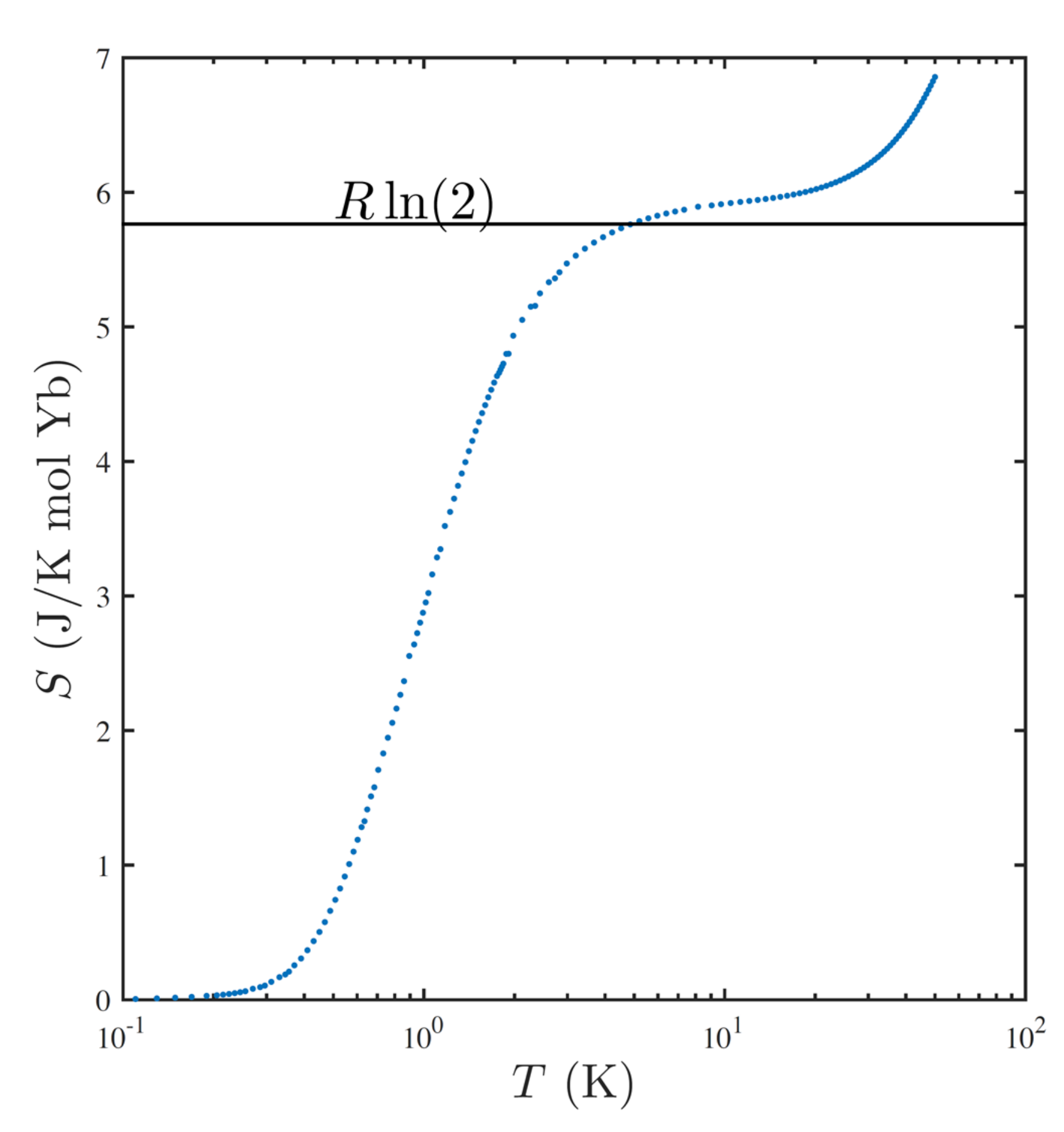}
\centering
\caption{Magnetic entropy extracted from the 0T specific heat measurement. The entropy reaches Rln2 per Yb between 50 mK and 5 K, indicating a low temperature effective spin-1/2.} \label{fig_entropy}
\end{figure}

\newpage
\bibliography{supplement}